\documentclass[a4paper,11pt]{article}

\usepackage{jheppub} 
                     
\usepackage{url}
\usepackage[compat=1.1.0]{tikz-feynman}
\usepackage{tikz}
\usetikzlibrary{calc,decorations.markings}
\usepackage{rotating}

\usepackage[T1]{fontenc} 
\usepackage[english]{babel}
\usepackage{amsmath,amssymb,euscript}
\usepackage{enumerate}
\usepackage{array}
\usepackage{graphicx}
\usepackage{listings, color}
\usepackage{calrsfs}
\usepackage{hyperref}
\usepackage{cleveref}
\usepackage{autonum}
\DeclareMathSymbol{*}{\mathbin}{symbols}{"01} 

\makeatletter
\newcommand{\thickhline}{%
	\noalign {\ifnum 0=`}\fi \hrule height 1pt
	\futurelet \reserved@a \@xhline
}
\newcolumntype{"}{@{\hskip\tabcolsep\vrule width 1pt\hskip\tabcolsep}}
\makeatother

\definecolor{darkblue}{rgb}{0,0,.6}
\definecolor{darkred}{rgb}{.6,0,0}
\definecolor{darkgreen}{rgb}{0,.6,0}
\definecolor{red}{rgb}{.98,0,0}

\newcommand{\eul}{\mathrm{e}}

\newcommand{\dd}{\mathrm{d}}

\newcommand{\Res}{\mathrm{Res}}

\newcommand{\fpartial}[2]{\frac{\partial #1}{\partial #2}}
\newcommand{\fnpartial}[3]{\frac{\partial^{#1} #2}{\partial{#3}^{#1}}}

\newcommand{\abs}[1]{\left|#1\right|}
\newcommand{\ket}[1]{\left\vert#1\right\rangle}

\newcommand{\braket}[2]{\left\langle #1\vert #2\right\rangle}
\newcommand{\melement}[3]{\left\langle #1\left\vert #2\right\vert #3\right\rangle}
\newcommand{\expect}[1]{\left\langle #1\right\rangle}

\newcommand{\Op}{\mathcal{O}}

\newcommand{\un}[1]{\underline{#1}}

\newcommand{\Nphi}{N_{\phi}}
\newcommand{\afrk}[2]{\mathfrak{a}^{(#1)}_{#2}}

\begin{document}
\begin{flushright}
	LMU-ASC 36/20\\
\end{flushright}

\title{\boldmath Loops in dS/CFT}

 \author{Till Heckelbacher}
 \author{and Ivo Sachs}
 \affiliation{Arnold-Sommerfeld-Center for Theoretical Physics, Ludwig-Maximilians-Universit\"at of Munich,\\
 Theresienstr. 37, D-80333 M\"unchen, Germany}




\emailAdd{till.heckelbacher@physik.lmu.de}
\emailAdd{ivo.sachs@physik.lmu.de}


\abstract{We consider the semi-classical expansion of the Bunch-Davies wavefunction with future boundary condition in position space for a real scalar field, conformally coupled to a classical de Sitter background in the expanding Poincar\'e patch with quartic selfinteraction. In the future boundary limit the wave function takes the form of the generating functional of a Euclidean conformal field theory for which we calculate the anomalous dimensions of the double trace deformations at one loop order using results obtained from Euclidean Anti de Sitter space. We find analytic expressions for some subleading twist operators and an algorithm to obtain expressions for general twist.}

\maketitle
\flushbottom

\section{Introduction}
The most convincing explanation for current observations of the dynamics in cosmology suggests that the geometry of our universe can be well approximated by de Sitter space-time in the very early inflationary stage and in the asymptotic future. To understand phenomena which originate from the early universe like structure formation and temperature fluctuations in the cosmological microwave background (CMB) \cite{Mukhanov:1981xt} it is therefore of interest to understand the behaviour of quantum fields in a de Sitter background.

Another motivation to study a scalar quantum field in a de Sitter geometry comes from the idea that there is a duality between the quantum theory in de Sitter space and a conformal field theory living at future infinity. This idea was developed in e.g. \cite{Witten:2001kn,Maldacena:2002vr,Strominger:2001pn,Anninos:2014lwa,Anninos:2011ui}. The main idea is that the wavefunction describing the Bunch-Davies vacuum at late times can be calculated from the partition function of a conformal field theory living in one less dimension. By comparing the wavefunction of de Sitter with the partition function of a QFT in Euclidean Anti de Sitter space (EAdS) we can see that they are related by a simple double analytic continuation as was described in \cite{Maldacena:2002vr}. Further work on this topic was also done in amongst others \cite{Ng:2012xp,Anninos:2013rza,Anninos:2017eib,Sleight:2019hfp,Konstantinidis:2016nio,Strominger:2001gp,Baumann:2020dch}.

The main goal of this work is to calculate contributions to the wave function up to second order in the perturbative expansion including loops and to use existing results from EAdS to learn more about the dual CFT. This approach is similar to work done in \cite{Anninos:2014lwa,Gorbenko:2019rza} where the focus is on infrared effects in de Sitter space. We use calculations in position space to make the connection to the conformal block expansion in the dual CFT explicit. We will calculate the loop corrections to the two point and four point function of a conformally coupled, real scalar field in the expanding Poincar\'e patch of de Sitter with a quartic interaction term. The corresponding calculations for EAdS have been done in \cite{Bertan:2018afl,Bertan:2018khc}. Interestingly we will find that the results for the anomalous dimensions of the double trace operators in dS are actually the same.

The difference between the calculations in AdS and dS is the treatment of the vacuum with respect to the background. In de Sitter the background metric is time dependent and there exists no global time-like Killing vector. Therefore the in- and out- vacua are not the same state and care has to be taken when calculating correlation functions (see \cite{Allen:1985ux,Fukuma:2013mx} for discussions of that topic).

To find the effects of quantum corrections in de Sitter or any other time dependent space time most of the work is usually done in the Schwinger-Keldysh formalism (see e.g. \cite{Akhmedov:2013vka}). In the present work we instead calculate the wave function of the Bunch-Davis vacuum by calculating the on-shell action of the scalar field in de Sitter and calculate quantum corrections by using the background field method \cite{Anninos:2014lwa}, which in turn can be obtained by a double analytic continuation of the results obtained for EAdS \cite{Bertan:2018afl,Bertan:2018khc}.

In particular, the quantum contributions to the two point function are all proportional to a massshift und the renormalization scheme can then be defined such that the physical mass is fixed by the scaling behaviour on the boundary and so no anomalous dimensions are picked up and no information about the CFT can be obtained from the two point function. So, in order to obtain anomalous dimensions at this order in perturbation theory one has to consider the four point function.

In section 4 we explore the dual conformal field theory and calculate the anomalous dimensions and conformal block coefficients. We show that they can be obtained from those in EAdS which where found in \cite{Bertan:2018afl}, which is one key result.\footnote{Mind however, that in \cite{Bertan:2018afl} a different sign for the interaction term was used} In addition we extend these calculations allowing us to obtain contributions to the anomalous dimensions up to spin $l=18$. This allows us to generalize the analytic expression for the anomalous dimensions of the leading twist operators at one loop order from \cite{Bertan:2018afl,Bertan:2018khc} to subleading twist operators.

Finally, we comment on potential implications of our results on the conjectured duality between a higher spin Vasiliev theory in the de Sitter bulk and an $Sp(N)$ vector model as described in \cite{Anninos:2011ui} in analogy to the $O(N)$ vector model in AdS.

\section{Classical de Sitter background}
In this work we are concerned with quantum fields in a classical de Sitter background, which has the following geometry.

De Sitter space is a maximally symmetric space with constant positive curvature i.e. the  Ricci scalar fulfills the condition $R>0$. It can be visualized as a hyperboloid:
\begin{align}
	-(X^0)^2+\sum\limits_{i=1}^4(X^i)^2=\ell^2
\end{align}
in a 5 dimensional ambient space with metric:
\begin{align}
	\dd s^2=\eta_{ij}\dd X^i\dd X^j
\end{align}
where $\eta_{ij}=\mathrm{diag}(-1,1,1,1,1)$ is the 5 dimensional Minkowski metric and $\ell$ is the de Sitter radius.

Now we can try to find local coordinates which obey the de Sitter conditions. We will use the so called Poincar\'e patch which is spatially flat and describes only on half of the whole de Sitter manifold. However it is the one which is most relevant in cosmology as it is believed to approximately describe the universe during the inflationary stage and in the assymptotic future.
In flat coordinates with Lorentzian signature the coordinate system is given by:
\begin{align}
	X^0&=\frac{\ell}{\sqrt{2}\eta}\left(1+\frac{\un{x}^2}{2}-\frac{\eta^2}{2}\right)\label{eq:dScoord}\\
	X^i&=\frac{x^i\ell}{\eta}\\
	X^4&=\frac{\ell}{\sqrt{2}\eta}\left(1-\frac{\un{x}^2}{2}+\frac{\eta^2}{2}\right)
\end{align}
where $\eta$ ranges from $-\infty$ to $0$ and the asymptotic future lies at $\eta=0$.

Then the induced metric is given by:
\begin{align}
	\dd s^2&=\frac{\ell^2}{\eta^2}(-\dd\eta^2+\dd x^2+\dd y^2+\dd z^2)
\end{align}
Now we can define a quantity
\begin{align}
	Z(X,Y)=&\frac{\eta_{ij}X^iY^j}{\ell^2}
\end{align}
which is clearly preserved under the de Sitter condition and is therefore a kinematical invariant of the de Sitter group which can be related to the geodesic distance. In the flat coordinates introduced above it is given by:
\begin{align}
	Z=&1+\frac{(\eta_x-\eta_y)^2-(\un{x}-\un{y})^2}{2\eta_x\eta_y}=\frac{\eta_x^2+\eta_y^2-(\un{x}-\un{y})^2}{2\eta_x\eta_y}\equiv\frac{1}{P_{xy}}\label{eq:hypdist}
\end{align}
$Z$ is called the hyperbolic distance and is related to the physical geodesic distance $d(x,y)$ by
\begin{align}
	Z(x,y)=\cos\frac{d(x,y)}{\ell}
\end{align}
We can see from \eqref{eq:hypdist} that for points with light-like separation we get $Z=1$ for $Z<1$ we have space-like separation and for $Z>1$ time-like separation.

We can also see that for values of $Z\leq-1$ the following condition has to be fulfilled:
\begin{align}
	(\eta_x+\eta_y)^2-(\un{x}-\un{y})^2\leq0
\end{align}
We see that this is the condition for one of the points being in the lightcone of the other point after making the replacement $\eta_y\rightarrow-\eta_y$. Comparing this to \eqref{eq:dScoord} we see that this transformation is nothing but going from the point $Y^i$ to the antipodal point $-Y^i$. As the points with positive $\eta$ are not covered by Poincar\'e coordinates we see that these values of $Z$ are out of reach. This will play an important role in the next section when we analyze the Green functions of a scalar field theory. 

\subsection{Classical solutions of the conformally coupled scalar field}
Now we can look at the main topic of this work which is a real scalar field theory with a quartic interaction. The action of this is given by:
\begin{equation}
	S[\phi]=\frac{\ell^2}{2}\int\limits_{-\infty}^0\frac{\dd^3x\dd\eta}{\eta^4}\left\{\eta^2(\partial_{\eta}\phi)^2-\eta^2(\nabla\phi)^2-m^2\ell^2\phi^2-\frac{\lambda \ell^2}{12}\phi^4\right\}\label{eq:action}
\end{equation}
Eventually we would like to solve this theory perturbatively in orders of $\lambda$. Therefore we start with the solution of the free theory.
The classical equation of motion of the free scalar field in the Poincare patch of de Sitter space is given by:
\begin{align}
	\eta^2\phi''(\eta,\un{x})-2\eta\phi'(\eta,\un{x})-\eta^2\Delta\phi(\eta,\un{x})+m^2\ell^2\phi(\eta,\un{x})=0\label{eq:eom}
\end{align}
Performing a partial Fourier transformation for the spatial part of $\phi(\eta,\un{x})=\int\dd^3k\phi_k (\eta)\eul^{-i\un{k}\un{x}}$ we can solve the equation of motion for each mode $\phi_k(\eta)$. The general solution to this equation is a superposition of the Bessel functions $Y_{\nu}(x)$ and $J_{\nu}(x)$. We fix the boundary conditions in the past so that the modes behave like free waves in the limit $\eta\rightarrow-\infty$. This defines the Bunch-Davies boundary condition (see e.g. \cite{Birrell:1982ix} for a detailed calculation). The mode function is then given by:
\begin{align}
	\phi_k(\eta)=&\frac{\sqrt{\pi}}{2\ell}\eta^{3/2}H_{\nu}^{(1)}(k\eta)=\frac{\sqrt{\pi}}{2\ell}\eta^{3/2}(J_{\nu}(k\eta)+iY_{\nu}(k\eta));\\
	\phi_k^{\ast}(\eta)=&\frac{\sqrt{\pi}}{2\ell}\eta^{3/2}H_{\nu}^{(2)}(k\eta)=\frac{\sqrt{\pi}}{2\ell}\eta^{3/2}(J_{\nu}(k\eta)-iY_{\nu}(k\eta));\quad 
	\text{with: }\nu=\sqrt{\frac94-m^2\ell^2}\label{eq:BD_modes}
\end{align}
Now we can construct Green functions for \eqref{eq:eom} from linear combinations of products of two mode functions. We are mainly interested in the massless conformally coupled case which corresponds to $m^2\ell^2=2$ and therefore $\nu=\frac12$. By the choice of the Green function we fix the boundary conditions on the future slice of $\eta\rightarrow0$. We will focus on three special cases which correspond to Dirichlet boundary conditions, Neumann boundary conditions and the superposition of both respectively:
\begin{align}
	iG_D(\eta,\eta';k)=&\frac{\pi(\eta\eta')^{3/2}}{2\ell^2}H_{1/2}^{(1)}(k\eta)J_{1/2}(k\eta)=-\frac{i\eta\eta'}{2k\ell^2}\eul^{ik\eta'}\sin(k\eta)\\
	iG_N(\eta,\eta';k)=&i\frac{\pi(\eta\eta')^{3/2}}{2\ell^2}H_{1/2}^{(1)}(k\eta)J_{1/2}(k\eta)=-\frac{\eta\eta'}{2k\ell^2}\eul^{ik\eta'}\cos(k\eta)\\
	iG_{BD}(\eta,\eta';k)=&\frac{\pi(\eta\eta')^{3/2}}{2\ell^2}H_{1/2}^{(1)}(k\eta)H_{1/2}^{(2)}(k\eta)=\frac{\eta\eta'}{2k\ell^2}\eul^{ik(\eta-\eta')}\label{eq:GBD}
\end{align}
The label "BD" for \eqref{eq:GBD} means that this is the Bunch Davies Green function. It is obtained by calculating the expectation value of the operator $\phi_k(\eta)\phi_{-k}(\eta')$ in the Bunch-Davies vacuum. On the other hand, it is easy to check that $G_D$ and $G_N$ satisfy the corresponding boundary conditions, i.e.
\begin{align}
	\lim\limits_{\eta\rightarrow0}\frac{1}{\eta}G_D(\eta,\eta';k)=0;\quad \lim\limits_{\eta\rightarrow0}\partial_{\eta}\left(\frac{1}{\eta}G_N(\eta,\eta';k)\right)=0
\end{align}
Now we can transform these functions back into position space to get:
\begin{align}
	iG_D(x,y)=&i\int\frac{\dd^3k}{(2\pi)^3}G_D(\eta,\eta';k)\eul^{-ik(\un{x}-\un{y})}=-\frac{1}{4\pi^2\ell^2}\frac{P^2}{1-P^2}\label{eq:GF_Dirichlet}\\
	iG_N(x,y)=&i\int\frac{\dd^3k}{(2\pi)^3}G_N(\eta,\eta';k)\eul^{-ik(\un{x}-\un{y})}=-\frac{1}{4\pi^2\ell^2}\frac{P}{1-P^2}\label{eq:GF_Neumann}\\
	iG_{BD}(x,y)=&i\int\frac{\dd^3k}{(2\pi)^3}G_{BD}(\eta,\eta';k)\eul^{-ik(\un{x}-\un{y})}=-\frac{1}{8\pi^2\ell^2}\frac{P}{1-P}\label{eq:GF_BD}
\end{align}
For the integral in this calculation to converge we have to give $\eta$ a small negative imaginary part $-i\varepsilon$ which we dropped for better readability.
Let us add a few comments:

By looking at \eqref{eq:GF_Dirichlet} and \eqref{eq:GF_Neumann} we see that $G_D$ and $G_N$ have two singularities at $P=\pm 1$. From the definition of $P$ we know that $P=1$ corresponds the points lying on the lightcone. In global de Sitter space $P=-1$ would correspond to the antipodal point but in the Poincar\'e patch this part is not covered as was discussed at the end of the previous section.

$G_{BD}$ is special in the sense, that it is the only Green function of the scalar field in dS which only has one pole at $P=1$, as can easily be seen from equation \eqref{eq:GF_BD}. We can rewrite $G_{BD}$ in the follwoing way:
\begin{align}
	iG_{BD}(x,y)=&-\frac{1}{8\pi^2\ell^2}\left(\frac{P}{1-P^2}+\frac{P^2}{1-P^2}\right)
\end{align}
Comparing this to \eqref{eq:GF_Dirichlet} and \eqref{eq:GF_Neumann} we see that the second term is the Green function with the Dirichlet boundary condition while the first term is the Green function with Neumann boundary condition.

It was shown (e.g. in \cite{Akhmedov:2019esv}) that the Dirichlet and the Neumann Green function cannot be written as a Wightman function between two same de Sitter invariant vacuum states. They are rather matrix elements between an in- and an out-state. In this case the in-state is the Bunch-Davies vacuum while for \eqref{eq:GF_Dirichlet} the out-state corresponds to a vacuum that is annihilated by the coefficient of the leading order term of the mode function at $\eta=0$ and for \eqref{eq:GF_Neumann} the out-state corresponds to a vacuum that is annihilated by the coefficient of the first subleading order term of the mode function at $\eta=0$. Therefore they act as a Dirichlet and Neumann boundary conditions respectively. See also \cite{Fukuma:2013mx} for a deeper analysis of this topic.

We are mainly interested in the Green functions with Dirichlet boundary condition. Here the connection to the quantities in Euclidean Anti-de-Sitter space becomes clear, as the Green function and the bulk to boundary propagator are the same as what you get from calculations in the Poincar\'e patch of EAdS when doing the double analytic continuation $z=-i\eta$ and $\ell_{AdS}=-i\ell$\cite{Anninos:2014lwa}. It is also the relevant Green function to calculate the wave function.

To do this we now want to solve the classical field equations of the interacting theory from \eqref{eq:action} perturbatively with Dirichlet boundary conditions, i.e. we fix the value of the field at some time $\eta=\epsilon$ to $\phi_k(\eta)=\phi_0(k)$:
\begin{align}
	\phi_k(\eta,\epsilon)=\left(\frac{\eta}{\epsilon}\right)^{3/2}\frac{H^{(1)}_{\nu}(k\eta)}{H^{(1)}_{\nu}(k\epsilon)}\phi_0(k)
\end{align}
Again we are mainly interested in the massless conformally coupled case which corresponds to $m^2\ell^2=2$ and therefore $\nu=\frac12$. The solution of the free field in position space with fixed value at $\eta=\epsilon$ is therefore:
\begin{align}
	\phi(\eta,\epsilon;\un{x})=&\int\frac{\dd^3k}{(2\pi)^3}\left(\frac{\eta}{\epsilon}\right)^{3/2}\frac{H^{(1)}_{1/2}(k\eta)}{H^{(1)}_{1/2}(k\epsilon)}\phi_0(k)\eul^{-i\un{k}\un{x}}
	=\int\dd^3\un{y}\int\frac{\dd^3k}{(2\pi)^3}\left(\frac{\eta}{\epsilon}\right)^{3/2}\frac{H^{(1)}_{1/2}(k\eta)}{H^{(1)}_{1/2}(k\epsilon)}\eul^{-i\un{k}(\un{x}-\un{y})}\phi_0(\un{y})\\
	=&\frac{1}{2\pi^2}\frac{\eta}{\epsilon}\int\dd^3\un{y}\frac{\phi(\un{y})}{\abs{\un{x}-\un{y}}}
	\int_{0}^{\infty}\dd k k\sin(k\abs{\un{x}-\un{y}})\eul^{ik(\epsilon-\eta)}\phi_0(\un{y})\\
	=&\int\dd^3\un{y}\underbrace{\frac{i}{\pi^2}
		\frac{\eta(\eta-\epsilon)}{\epsilon((\eta-\epsilon)^2-\abs{\un{x}-\un{y}}^2-i\varepsilon)}}_{K(\un{y},x)}\phi_0(\un{y})
\end{align}
We will call the integral kernel $K(\un{y},x)$ the bulk to boundary propagator in correspondence to the AdS/CFT convention. Eventually we will be interested in the values of $\phi$ on the future boundary which corresponds to $\epsilon\rightarrow 0$ and the bulk to boundary propagator becomes:
\begin{align}
	K(\un{y},x)=\frac{i}{\pi^2\epsilon}\frac{\eta^2}{\left(\eta^2-\abs{\un{x}-\un{y}}^2-i\varepsilon\right)^2}\equiv\frac{i}{\pi^2}\bar{P}_{\un{y},x}
\end{align}
Now we want to calculate the solution to the interacting field perturbatively. To first order the solution to the equation of motion is given by:
\begin{align}
	\phi(\un{x},\eta)=&\underbrace{\int\dd^3\un{y}K(\un{y},x)\phi_0(\un{y})}_{:=\phi_1(x)}-\frac{\lambda}{3!}\int\dd^4y\sqrt{g(y)}G_D(x,y)\phi_1^3(y)
\end{align}
Where $G_D(x,y)$ is the Green function of the free equation of motion also fulfilling the Dirichlet boundary condition $G_D(x,y)\vert_{\eta_x=\epsilon}=0$ at finite $\epsilon$	In the limit $\eta\rightarrow0$ it reduces to \eqref{eq:GF_Dirichlet} \cite{Anninos:2014lwa}.
With these results we can now calculate the on-shell action to first order in the coupling constant.

To vizualize the calculation the Witten diagrams developed for AdS/CFT can be applied to this calculation in a slightly modified way:
\begin{enumerate}
	\item The \textit{bulk to boundary propagator} $K(\un{y},x)$ with point $\un{y}$ at future infinity and $x$ at final time is represented by:
	\begin{align}
		\begin{tikzpicture}[baseline=(base)]
			\begin{feynman}[inline=(a)]
				\vertex (a);
				\vertex [right=1.3cm of a] (b);
				\tikzfeynmanset{every vertex=dot}
				\vertex [right=0.1cm of a, label=90:$\un{y}$] (y);
				\vertex [below right=0.5cm and 0.5cm of y, label=0:$x$] (x);
				\tikzfeynmanset{every vertex={empty dot,minimum size=0mm}}
				\vertex [below right =0.2cm and 0.5cm of y] (base);
				\diagram* {
					(a)--[double](y)--[double](b),
					(y)--[out=280, in=180, min distance=0.3cm](x),
				};
			\end{feynman}
		\end{tikzpicture}\label{eq:witten_bulkbulk}
	\end{align}
	
	\item The \textit{bulk to bulk propagator} $iG(x,y)$ with points $x$ and $y$ at final time:
	\begin{align}
		\begin{tikzpicture}[baseline=(base)]
			\begin{feynman}[inline=(a)]
				\vertex (a);
				\vertex [right=1.3cm of a] (b);
				\tikzfeynmanset{every vertex=dot}
				\vertex [below right=0.3 cm and 0.4cm of a, label=180:$y$] (y);
				\vertex [below right=0.3cm and 0.9cm of a, label=0:$x$] (x);
				\tikzfeynmanset{every vertex={empty dot,minimum size=0mm}}
				\vertex [below right =0.2cm and 0.5cm of y] (base);
				\diagram* {
					(a)--[double](b),
					(y)--(x),
				};
			\end{feynman}
		\end{tikzpicture}\label{eq:witten_bulkboundary}
	\end{align}
\end{enumerate}

\subsection{The Bunch Davies vacuum and its wave function}
When we were solving the equations of motion of the scalar field in a de Sitter universe we saw in the previous section at \eqref{eq:BD_modes} that there is a special solution for the mode functions which behaves like the modes in flat space for $k\eta\rightarrow -\infty$. The vacuum state which is annihilated by these modes is the Bunch-Davies vacuum.

In this work we would like to quantize our theory by calculating the wave function of the Bunch-Davies vacuum as a path integral \cite{Maldacena:2002vr}. This works analogously to quantum mechanics. Here the wave function of a state $\ket{\psi,t}$ can be represented by $\psi(q)=\braket{q}{\psi,t}$ where $\ket{q}$ is the eigenstate of an operator, in this case the position operator $q$. The propagator from a position eigenstate with eigenvalue $q_i$ at time $t_i$ to the position $q_f$ at time $t_f$ can be written in the following way:
\begin{align}
	\braket{q_f,t_f}{q_i,t_i}=&\melement{q_f,t_f}{\eul^{iH(t_f-t_i)}}{q_i,t_f}
\end{align}
If we set the final time and postition $t_f=0$ and $q_f=q$ and the initial time and position to $t_i=t$ and $q_i=0$ this expression becomes \cite{Hartle:1983ai}:
\begin{align}
	\braket{q,0}{0,t}=&\melement{q,0}{\eul^{-iHt}}{0,0}=\int\limits_{\substack{q(0)=q\\q(t)=0}}\mathcal{D}q\eul^{iS[q]}
\end{align}
Considering a bound particle we can write this in a different way by inserting a complete set of energy eigenstates with eigenvalues $E_n$ as:
\begin{align}
	\braket{q,0}{0,t}=&\melement{q,0}{\eul^{-iHt}}{0,0}=\sum\limits_n\eul^{-iE_nt}\psi_n(q)\psi^{\ast}_n(0)
	\stackrel{t\rightarrow-\infty(1+i\varepsilon)}{\longrightarrow}\eul^{iE_0\infty(1+i\varepsilon)}\psi_0(q)\psi^{\ast}_0(0)
\end{align}
The limit $t\rightarrow-\infty(1+i\varepsilon)$ projects out the vacuum state, as $E_0$ is the smallest value in the sum. Therefore the wavefunction of the vacuum $\psi_0(q)$ at time $t=0$ is given by:
\begin{align}
	\psi_0(q)=&\lim\limits_{t\rightarrow-\infty(1+i\varepsilon)}\frac{1}{\psi^{\ast}_0(0)}\int\limits_{\substack{q(0)=q\\q(t)=0}}\mathcal{D}q\eul^{iS[q]}
\end{align}
This analysis can be translated into quantum field theory in de Sitter space. The wave function of state $\ket{\Psi}$ is given by $\Psi[\eta,\phi(x)]=\braket{\phi(\eta,x)}{\Psi}$ where $\ket{\phi(\eta,x)}$ is an eigenstate of the field operator at a fixed time slice $\eta$. This means that $\Psi[\eta,\phi(x)]$ is the amplitude to find the field configuration $\phi(x)$ at time $\eta$. To find the wave function of the Bunch-Davies vacuum we have to find a similar way to project out the vacuum. As there is no globally defined Fock space we cannot make the same reasoning as for the quantum mechanical example, where we used the fact that the vacuum corresponds to the lowest energy eigenvalue. However we can use a different property of the Bunch-Davies vacuum. It is defined by its behaviour at $\eta\rightarrow-\infty$ which corresponds to a free wave in flat space, i.e. $\propto\eul^{-ik\eta}$. In the limit $\eta\rightarrow-\infty(1+i\varepsilon)$ this vanishes. So we can require this as the boundary condition and write the wave function of the Bunch-Davies vacuum at time $\eta$ as the path integral:	
\begin{align}
	\Psi[\eta,\phi_0(\un{x})]=&\lim\limits_{\eta'\rightarrow-\infty(1+i\varepsilon)}\int\limits_{\substack{\phi(\eta,\un{x})=\phi_0(\un{x})\\\phi(\eta',\un{x})=0}}\mathcal{D}\phi\eul^{iS[\phi]}\label{eq:BD_wavefunction}
\end{align}
It was discussed e.g. in \cite{Maldacena:2002vr,Anninos:2014lwa} that there is a straightforward relation between the Bunch-Davies wave function with future boundary condition \eqref{eq:BD_wavefunction} and the partition function of a scalar field in the Poincar\'e patch of Euclidean Anti de Sitter space:
\begin{align}
	\Psi[\eta,\phi_0(\un{x})]=\left.Z_{EAdS}[i z,\phi_0(\un{x})]\right\vert_{\ell=-i\ell_{AdS}}=\int\mathcal{D}\phi\eul^{-S_{EAdS}[\phi]}\label{eq:partitionfunction}
\end{align}
So in analogy with the AdS/CFT correspondence we expect there to be a dS/CFT correspondence where the correlation functions of the CFT are generated by functional derivatives of the Bunch-Davies wavefunction with respect to the values of the field at the boundary $\eta\rightarrow 0$:
\begin{align}
	\expect{\Op(\un{x}_1)\Op(\un{x}_2)...\Op(\un{x}_n)}=&\lim\limits_{\eta\rightarrow 0}
	\frac{\delta^n}{\delta\phi_0(\un{x}_1)\delta\phi_0(\un{x}_2)...\delta\phi_0(\un{x}_n)}\Psi[\eta,\phi_0(\un{x})]\label{eq:npoint}
\end{align}
In the next section we will use direct calculations in de Sitter and use results from EAdS from \cite{Bertan:2018afl} to find information about this expected CFT.

\section{Semi-classical expansion of the Wave function}
Using the results from the previous section for the Green function and the Bunch-Davies wave function we now would like to make a semiclassical expansion of the wave function to calculate the contributions to the two point and four point function of the conformal field theory up to second order in the coupling constant.

Mathematically this corresponds to a saddle point approximation around the classical action which contributes mostly to the wave function.

\subsection{Tree-level contributions}
The tree-level contributions to the wave function of the scalar field are given by the classical onshell action. They are obtained by expanding the field to first order in the coupling constant and plugging it into the equation of motion. Doing partial integration of the action and using the equation of motion we get:
\begin{align}
	S_{on-shell}=&\frac12\int\dd^3\un{x}\frac{\ell^2}{\eta^2}\phi_1(x)\partial_{\eta}\phi_1(x)\vert_{\eta=\epsilon}
	-\frac{\lambda}{4!}\int\dd^4x\sqrt{g(x)}\phi_1^4(x)
\end{align}
\subsubsection{Two point function}
The first term generates the two point function and is given by:
\begin{align}
	\left.\frac12\int\dd^3\un{x}\frac{\ell^2}{\eta^2}\phi_1(x)\partial_{\eta}\phi_1(x)\right\vert_{\eta=\epsilon}=&
	\left.\frac12\int\dd^3\un{x}\frac{\ell^2}{\eta^2}\phi_1(x)\partial_{\eta}\int\dd^3\un{y}\frac{i}{\pi^2}
	\frac{\eta(\eta-\epsilon)}{\epsilon\left((\eta-\epsilon)^2-\abs{\un{x}-\un{y}}^2\right)^2}\phi_0(\un{y})\right\vert_{\eta=\epsilon}\\
	=&\frac{i\Nphi}{2}\int\dd^3x\dd^3y\frac{\phi_0(x)\phi_0(y)}{\abs{x-y}^4};\quad\text{with $\Nphi=\frac{\ell^2}{\pi^2\epsilon^2}$}\label{eq:on_shell_2pnt}
\end{align}
This coincides with the result from \cite{Anninos:2014lwa} after Fourier transformation. It is also just the result from calculations done in AdS after the double analytical continuation \cite{Muck:1998rr}.
Now the tree level contribution to the wave function is given by:
\begin{align}
	\Psi[\phi_0,\epsilon]&=\eul^{i S_{on-shell}}\\
	S_{on-shell}&=\frac{i\Nphi}{2}\int\dd^3x\dd^3y\frac{\phi_0(x)\phi_0(y)}{\abs{x-y}^4}\\
	&-\frac{\lambda}{4!\epsilon^4\pi^8}\iint\dd^4y\sqrt{g(y)}\prod\limits_{i=1}^4\dd^3x_i\frac{\eta^8\phi_0(x_i)}{(\eta^2-\abs{x_i-y}^2)^2}\label{eq:onshellaction}
\end{align}
The two point function can now be calculated with the help of equation \eqref{eq:npoint} and written in a diagramatic form with \eqref{eq:witten_bulkbulk} and \eqref{eq:witten_bulkboundary}:
\begin{align}
	\expect{\Op(x_1)\Op(x_2)}=	\frac{\delta^2}{\delta\phi_0(\un{x}_1)\delta\phi_0(\un{x}_2)}\eul^{iS_{on-shell}}=\begin{tikzpicture}[baseline=(base)]
		\begin{feynman}[inline=(a)]
			\vertex (a);
			\vertex [right=1.3cm of a] (b);
			\tikzfeynmanset{every vertex=dot}
			\vertex [right=0.1cm of a, label=90:$x_1$] (x1);
			\vertex [right=1cm of x1, label=90:$x_2$] (x2);
			\tikzfeynmanset{every vertex={empty dot,minimum size=0mm}}
			\vertex [below=0.2cm of x1] (base);
			\diagram* {
				(a)--[double](x1)--[double](x2)--[double](b),
				(x1)--[out=280, in=260, min distance=0.6cm](x2)
			};
		\end{feynman}
	\end{tikzpicture}=-\frac{\Nphi}{\abs{\un{x}_1-\un{x}_2}^4}
\end{align}

\subsubsection{Four point function}
The next order tree level contribution we can construct from the on shell action is the connected part of the four point function which is generated by the second term in \eqref{eq:onshellaction}. In diagramatic form it is given by:
\begin{align}
	\begin{tikzpicture}[baseline=(base)]
		\begin{feynman}[inline=(a)]
			\vertex (a);
			\vertex [right=3.7cm of a] (b);
			\tikzfeynmanset{every vertex=dot}
			\vertex [right=0.3cm of a, label=90:$x_1$] (x1);
			\vertex [below right=1cm and 1.55cm of x1, label=270:$x$] (x);
			\vertex [right=of x1, label=90:$x_2$] (x2);
			\vertex [right=of x2, label=90:$x_3$] (x3);
			\vertex [right=of x3, label=90:$x_4$] (x4);
			\tikzfeynmanset{every vertex={empty dot,minimum size=0mm}}
			\vertex [above=0.5cm of x] (base);
			\diagram* {
				(a)--[double](x1)--[double](x2)--[double](x3)--[double](x4)--[double](b),
				(x1)--(x)--(x2),
				(x3)--(x)--(x4),
			};
		\end{feynman}
	\end{tikzpicture}=&\frac{\Nphi^2}{\pi^4}M_4(\un{x}_1,\un{x}_2,\un{x}_3,\un{x}_4)
\end{align}
where $M_4$ is given by the integral:
\begin{align}
	M_4&=\int\limits_{-\infty}^0\dd\eta\dd^3 x\eta^4\frac{(\eta^2-(\un{x}_3-\un{x})^2-i\epsilon)^{-2}(\eta^2-(\un{x}_4-\un{x})^2-i\epsilon)^{-2}}{(\eta^2-(\un{x}_1-\un{x})^2-i\epsilon)^2(\eta^2-(\un{x}_2-\un{x})^2-i\epsilon)^2}
	\label{eq:4point_int}
\end{align}
Similarly to the calculation in EAdS \cite{Bertan:2018afl,Muck:1998rr} we can rewrite this integral in terms of Schwinger parameters by using the relation:
\begin{align}
	\frac{1}{A^2}&=\int\limits_0^{\infty}\dd\alpha\alpha\eul^{-i\alpha A}\qquad\text{with: }\mathfrak{Im}(A)<0
\end{align}
Therefore equation \eqref{eq:4point_int} can be reformulated in the following way:
\begin{align}
	M_4=&\int\limits_{-\infty}^{0}\dd\eta\dd^3 x\eta^4\int\limits_0^{\infty}\prod\limits_{i=1}^{4}\dd\alpha_i\alpha_i\eul^{-i\alpha_i(\eta^2-(\un{x}_i-\un{x})^2-i\epsilon)}\label{eq:4pointSchwinger}
\end{align}
Now the integration over $x$ and $\eta$ can be performed by rearranging the exponent to do a Gaussian integration (see Appendix A). The result is given by:
\begin{align}
	M_4=&\frac{-3i\pi^2}{8}\int\limits_0^{\infty}\prod_{i=1}^{4}\dd\alpha_i\alpha_i\frac{1}{\beta^4}\exp\left(-\frac{1}{i\beta}\left(\sum_{i<j=1}^4\alpha_i\alpha_j x_{ij}^2\right)\right)\label{eq:schwinger_par}
\end{align}
The exponential function can be written as an integral over the gamma function known as Mellin tranformation:
\begin{align}
	\eul^{-x}&=\frac{1}{2\pi i}\int\limits_{c-i\infty}^{c+i\infty}\Gamma(s)x^{-s}\dd s
\end{align}
Using this tranformation we can perform the integration over the Schwinger parameters in \eqref{eq:schwinger_par} (see Appendix A) to get the final result:
\begin{align}
	M_4=&\frac{-3i\pi^2}{4}\frac{1}{(x_{12}x_{34})^4}\left(\frac{u}{v}\right)^2\sum_{n,m=0}^{\infty}\frac{(1-v^{-1})^n(u/v)^m}{n!(m!)^2}
	\frac{\Gamma(2+m)^2\Gamma(2+m+n)^2}{\Gamma(4+2m+n)}\\
	&\times\left(\psi(1+m)-\frac12\log\left(\frac{u}{v}\right)-\psi(2+m)+\psi(4+2m+n)-\psi(2+m+n)\right)\\
	=&-i\frac{3\pi^2}{4}\frac{1}{(x_{12}x_{34})^4}\left(\frac{u}{v}\right)^2L_0\label{eq:m4}\\
	L_0=&\sum_{n,m=0}^{\infty}\frac{(1-v^{-1})^n(u/v)^m}{n!(m!)^2}
	\frac{\Gamma(2+m)^2\Gamma(2+m+n)^2}{\Gamma(4+2m+n)}\\
	&\times\left(\psi(1+m)-\frac12\log\left(\frac{u}{v}\right)-\psi(2+m)+\psi(4+2m+n)-\psi(2+m+n)\right)
\end{align}
Here the conformal invariants $u$ and $v$ were introduced which are given by:
\begin{align}
	u=\frac{x_{12}^2x_{34}^2}{x_{13}^2x_{24}^2};\quad v=\frac{x_{14}^2x_{23}^2}{x_{13}^2x_{24}^2}\quad
	\text{with: }x_{ij}=\abs{\un{x}_i-\un{x}_j}
\end{align}
Note that the result \eqref{eq:m4} corresponds to an expansion around $u\rightarrow0$ and $v\rightarrow1$. This is equivalent to an expansion around $\un{x}_1\rightarrow\un{x}_2$ and $\un{x}_3\rightarrow\un{x}_4$.\\
Equation \eqref{eq:m4} has the same form as the one already obtained for AdS. The only difference is the prefactor which differs from the AdS case by $-i$.

With this result we can now write the tree level on-shell action to first order in $\lambda$ as:
\begin{align}
	S_{on-shell}&=\frac{i\Nphi}{2}\int\dd^3x\dd^3y\frac{\phi_0(x)\phi_0(y)}{\abs{x-y}^4}-\frac{\lambda\Nphi^2}{4!\pi^4}\int\prod\limits_{i=1}^4\dd^3\un{x}_i\phi_0(\un{x}_i)M_4\\
	&=\frac{i\Nphi}{2}\int\dd^3x\dd^3y\frac{\phi_0(x)\phi_0(y)}{\abs{x-y}^4}+\frac{1}{4!}\frac{3i\lambda N_{\phi}^2}{4\pi^2}\int\prod\limits_{i=1}^4\dd^3\un{x}_i\phi_0(\un{x}_i)\frac{1}{(x_{12}x_{34})^4}\left(\frac{u}{v}\right)^2L_0\\
\end{align}
The four point function is given by the functional derivative of the wave function with respect to the field values on the boundary. They can be represented in terms of Witten diagrams in the following way:
\begin{align}
	\expect{\Op(x_1)\Op(x_2)\Op(x_3)\Op(x_4)}=&\frac{\delta^4}{\delta\phi(x_1)\delta\phi(x_2)\delta\phi(x_3)\delta\phi(x_4)}\eul^{iS_{on-shell}}\\
	=&\begin{tikzpicture}[baseline=(base)]
		\begin{feynman}[inline=(base)]
			\vertex (a);
			\vertex [right=1.7cm of a] (b);
			\tikzfeynmanset{every vertex=dot}
			\vertex [right=0.1cm of a, label=90:$x_1$] (x1);
			\vertex [right=0.5cm of x1, label=90:$x_2$] (x2);
			\vertex [right=0.4cm of x2, label=90:$x_3$] (x3);
			\vertex [right=0.5cm of x3, label=90:$x_4$] (x4);
			\tikzfeynmanset{every vertex={empty dot,minimum size=0mm}}
			\vertex [below=0.2cm of x1] (base);
			\diagram* {
				(a)--[double](x1)--[double](x2)--[double](x3)--[double](x4)--[double](b),
				(x1)--[out=280, in=260, min distance=0.3cm](x2),
				(x3)--[out=280, in=260, min distance=0.3cm](x4),
			};
		\end{feynman}
	\end{tikzpicture}+\begin{tikzpicture}[baseline=(base)]
		\begin{feynman}[inline=(base)]
			\vertex (a);
			\vertex [right=1.7cm of a] (b);
			\tikzfeynmanset{every vertex=dot}
			\vertex [right=0.1cm of a, label=90:$x_1$] (x1);
			\vertex [right=0.4cm of x1, label=90:$x_2$] (x2);
			\vertex [right=0.6cm of x2, label=90:$x_3$] (x3);
			\vertex [right=0.4cm of x3, label=90:$x_4$] (x4);
			\tikzfeynmanset{every vertex={empty dot,minimum size=0mm}}
			\vertex [below=0.2cm of x1] (base);
			\diagram* {
				(a)--[double](x1)--[double](x2)--[double](x3)--[double](x4)--[double](b),
				(x1)--[out=280, in=260, min distance=0.7cm](x4),
				(x2)--[out=280, in=260, min distance=0.3cm](x3),
			};
		\end{feynman}
	\end{tikzpicture}+\begin{tikzpicture}[baseline=(base)]
		\begin{feynman}[inline=(base)]
			\vertex (a);
			\vertex [right=1.7cm of a] (b);
			\tikzfeynmanset{every vertex=dot}
			\vertex [right=0.1cm of a, label=90:$x_1$] (x1);
			\vertex [right=0.4cm of x1, label=90:$x_2$] (x2);
			\vertex [right=0.6cm of x2, label=90:$x_3$] (x3);
			\vertex [right=0.4cm of x3, label=90:$x_4$] (x4);
			\tikzfeynmanset{every vertex={empty dot,minimum size=0mm}}
			\vertex [below=0.2cm of x1] (base);
			\diagram* {
				(a)--[double](x1)--[double](x2)--[double](x3)--[double](x4)--[double](b),
				(x1)--[out=280, in=260, min distance=0.5cm](x3),
				(x2)--[out=280, in=260, min distance=0.5cm](x4),
			};
		\end{feynman}
	\end{tikzpicture}\\
	&-i\lambda 	\begin{tikzpicture}[baseline=(base)]
		\begin{feynman}[inline=(a)]
			\vertex (a);
			\vertex [right=1.8cm of a] (b);
			\tikzfeynmanset{every vertex=dot}
			\vertex [right=0.1cm of a, label=90:$x_1$] (x1);
			\vertex [below right=0.4cm and 0.75cm of x1, label=270:$x$] (x);
			\vertex [right=0.5cm of x1, label=90:$x_2$] (x2);
			\vertex [right=0.5cm of x2, label=90:$x_3$] (x3);
			\vertex [right=0.5cm of x3, label=90:$x_4$] (x4);
			\tikzfeynmanset{every vertex={empty dot,minimum size=0mm}}
			\vertex [below=0.2cm of x1] (base);
			\diagram* {
				(a)--[double](x1)--[double](x2)--[double](x3)--[double](x4)--[double](b),
				(x1)--(x)--(x2),
				(x3)--(x)--(x4),
			};
		\end{feynman}
	\end{tikzpicture}+\Op(\lambda^2)\\
	=&\frac{N_{\phi}^2}{x_{12}^4x_{34}^4}\left(1+u^2+\frac{u^2}{v^2}-\frac{3\lambda}{4\pi^2}\left(\frac{u}{v}\right)^2L_0\right)+\Op(\lambda^2)\label{eq:4pointfirstorder}
\end{align}
All higher orders in $\lambda$ are beyond the tree level and will be calculated in the next section using the background field method. The result \eqref{eq:4pointfirstorder} is exactly the same as the one obtained from EAdS \cite{Bertan:2018afl,Muck:1998rr}. The difference in the sign of the prefactor of the two point function does not affect the four point function.

\subsection{Quantum corrections}
The wave function of the full quantum theory is given by the path integral over all field configurations with Bunch-Davies boundary conditions in the past and fixed boundary conditions in the future:
\begin{align}
	\Psi[\phi(\un{x},\eta)]\vert_{\eta=\epsilon}&=\int\mathcal{D}\phi\eul^{iS[\phi]}=\eul^{i\Gamma[\phi_0]}
\end{align}
$\Gamma[\phi_0]$ is the effective action and depends on the boundary values of the field at $\phi_0(\un{x})=\phi(\un{x},\epsilon)$. We will calculate this effective action perturbatively in loop orders of quantum corrections by using the background field method. This means we write our field as $\phi(\un{x},\eta)=\varphi(\un{x},\eta)+\chi(\un{x},\eta)$ where $\varphi(\un{x},\eta)$ satisfies the classical equations of motion and $\chi$ accounts for the quantum fluctuations. $\chi$ is choosen such that it vanishes on the boundary. If we plug this ansatz into the action we get the following result:
\begin{align}
	S[\phi]=&\int\dd^4 x\sqrt{g}\left\{-\frac12\partial_{\mu}\phi\partial^{\mu}\phi-\frac12m^2\phi^2-\frac{\lambda}{4!}\phi^4\right\}\\
	=&S[\varphi]+\int\dd^4x\sqrt{g}\chi\left\{\square\varphi -m^2\varphi-\frac{\lambda}{3!}\varphi^3\right\}+
	\int\dd^4x\sqrt{g}\left\{-\frac12\partial_{\mu}\chi\partial^{\mu}\chi-\frac12 m^2\chi^2\right\}\\
	&+\lambda\int\dd^4x\sqrt{g}\left\{-\frac14\varphi^2\chi^2-\frac16\varphi\chi^3-\frac{1}{4!}\chi^4\right\}
\end{align}
The first term corresponds to the classical on shell action which was calculated in the previous section. It generates the tree-level Witten diagrams. The second term in the expansion vanishes as $\varphi$ obeys the classical equations of motion. Therefore only the two remaining parts depend on $\chi$ and we are going to call the first one $S_0[\chi]$ as it corresponds to the free part of the action whereas the last term is called $S_{int}$ as it generates the interaction terms.

Now we can plug this action into the path integral for the wave function. As the classical field is completely fixed by the equations of motion and the boundary conditions the only path integral we have to perform is over the quantum fluctuations $\chi$:
\begin{align}
	\Psi[\phi(\un{x},\eta)]\vert_{\eta=\epsilon}&=\eul^{i\Gamma[\phi_0]}=\eul^{iS_{on-shell}[\varphi]}\int\mathcal{D}\chi\eul^{iS_0[\chi]+iS_{int}[\varphi,\chi]}
\end{align}
To make a semiclassical expansion up to second loop order, we expand $\eul^{iS_{int}[\varphi,\chi]}$ up to second order in the coupling constant $\lambda$. For better readability we write $\varphi(x)=\varphi_x$:
\begin{align}
	\eul^{iS_{int}[\varphi,\chi]}=&1-i\lambda\int\dd^4 x\sqrt{g(x)}\left\{\frac14\varphi_x^2\chi_x^2+\frac16\varphi_x\chi_x^3+\frac{1}{4!}\chi_x^4\right\}\\
	&-\frac{\lambda^2}{2}\iint\dd^4x\dd^4y\sqrt{g(x)g(y)}\left\{\frac{1}{16}\varphi_x^2\varphi_y^2\chi_x^2\chi_y^2+\frac{1}{4!2}\varphi_x^2\chi_x^2\chi_y^4+\frac{1}{36}\varphi_x\varphi_y\chi_x^3\chi^3_y\right.\\
	&\left.+\frac{1}{4!3}\varphi_x\chi^3_x\chi^4_y+\frac{1}{12}\varphi_x^2\varphi_y\chi_x^2\chi_y^3+
	\frac{1}{(4!)^2}\chi_x^4\chi_y^4\right\}+\Op(\lambda^3)
\end{align}
The second last two terms in the $\lambda^2$ integral have an odd number of $\chi$ insertions and therefore vanish when performing the path integral. The last term just gives a contribution in the bulk and is independent of the classical part of the field $\varphi$. Therefore the functional derivatives with respect to the fields on the boundary vanish and the term does not contribute to the conformal correlation functions \eqref{eq:npoint}. The same is true for the last two terms in the $i\lambda$ contribution. Now we can calculate the effective action perturbatively by using Wick's theorem to calculate the path integral over $\chi$:
\begin{align}
	\Psi[\phi(\un{x},\eta)]\vert_{\eta=\epsilon}=&\eul^{iS_{on-shell}[\varphi]}\left(1-\frac{i\lambda}{4}\int\dd^4x\sqrt{g(x)}\varphi_x^2iG(x,x)\right.\\
	&-\frac{\lambda^2}{2}\iint\dd^4x\dd^4y\sqrt{g(x)g(y)}\left\{\frac{1}{16}\varphi_x^2\varphi_y^2iG(x,x)iG(y,y)
	+\frac{1}{8}\varphi_x^2\varphi_y^2i^2G^2(x,y)\right.\\
	&\left.\left.+\frac{1}{4}\varphi_x^2i^2G^2(x,y)iG(y,y)+\frac16\varphi_x\varphi_yi^3G^3(x,y)\right.\right.\\
	&\left.\left.+\frac14\varphi_x\varphi_yiG(x,x)iG(x,y)iG(y,y)\right\}\right)
\end{align}
Plugging in the tree-level solution for $\varphi$ we can regroup the terms in the effective action to give the contributions to the 2 and 4 point functions. To simplify the notation we write $\int\dd^4x\sqrt{g(x)}=\int\dd^4\tilde{x}$ and $G(x,y)=G_{xy}$:
\begin{align}
	\Psi[\phi(\un{x},\eta)]\vert_{\eta=\epsilon}=&\eul^{i\Gamma[\phi_0]}
	=\eul^{iS_{on-shell}}\left(1-
	\iint\dd^3\un{x}_1\dd^3\un{x}_2\phi_0(\un{x}_1)\phi_0(\un{x}_2)\left[\frac{i\lambda}{4}\int\dd^4\tilde{x}iG_{xx}K_{x\un{x}_1}K_{x\un{x}_2}\right.\right.\\
	&+\lambda^2\iint\dd^4\tilde{x}\dd^4\tilde{y}\left\{\frac18i^2G^2_{xy}iG_{yy}K_{x\un{x}_1}K_{x\un{x}_2}
	+\frac{1}{12}i^3G^3_{xy}K_{x\un{x}_1}K_{y\un{x}_2}\right.\\
	&\left.\left.+\frac18iG_{xx}iG_{xy}iG_{yy}K_{x\un{x}_1}K_{x\un{x}_2}\right\}\right]\\
	&-\lambda^2\iiiint\prod\limits_{i=1}^4\dd^3\un{x}_i\phi_0(\un{x}_i)\left[\iint\dd^4\tilde{x}\dd^4\tilde{y}\left\{
	\frac{1}{3!2}iG_{xx}iG_{xy}K_{x\un{x}_1}K_{x\un{x}_2}K_{y\un{x}_3}K_{y\un{x}_4}\right.\right.\\
	&\left.\left.\left.+\frac{1}{32}iG_{xx}iG_{yy}K_{x\un{x}_1}K_{x\un{x}_2}K_{y\un{x}_3}K_{y\un{x}_4}
	+\frac{1}{16}i^2G^2_{xy}K_{x\un{x}_1}K_{x\un{x}_2}K_{y\un{x}_3}K_{y\un{x}_4}\right\}\right]+\Op(\lambda^3)\right)
\end{align}
By functional differentiating we can now calculate the two point and and four point function of the dual CFT up to second order in $\lambda$. In what follows we focus on these correlators rather than the wave function since they contain the relevant information about the CFT. We will write them in terms of the corresponding Witten diagrams:
\begin{align}
	\expect{\Op(\un{x}_1)\Op(\un{x}_2)}=\frac{\delta^2\Psi[\phi_0]}{\delta\phi_0(\un{x}_1)\delta\phi_0(\un{x}_2)}=&\begin{tikzpicture}[baseline=(base)]
		\begin{feynman}[inline=(a)]
			\vertex (a);
			\vertex [right=1.3cm of a] (b);
			\tikzfeynmanset{every vertex=dot}
			\vertex [right=0.1cm of a, label=90:$x_1$] (x1);
			\vertex [right=1cm of x1, label=90:$x_2$] (x2);
			\tikzfeynmanset{every vertex={empty dot,minimum size=0mm}}
			\vertex [below=0.2cm of x1] (base);
			\diagram* {
				(a)--[double](x1)--[double](x2)--[double](b),
				(x1)--[out=280, in=260, min distance=0.6cm](x2)
			};
		\end{feynman}
	\end{tikzpicture}
	-\frac{i\lambda}{2}	\begin{tikzpicture}[baseline=(y)]
		\begin{feynman}[inline=(a)]
			\vertex (a);
			\vertex [right=1.3cm of a] (b);
			\tikzfeynmanset{every vertex=dot}
			\vertex [right=0.1cm of a, label=90:$x_1$] (x1);
			\vertex [right=1cm of x1, label=90:$x_2$] (x2);
			\vertex [below right=0.5cm and 0.5cm of x1] (x);
			\tikzfeynmanset{every vertex={empty dot,minimum size=0mm}}
			\vertex [below right =0.2cm and 0.5cm of x1] (y);
			\diagram* {
				(a)--[double](x1)--[double](x2)--[double](b),
				(x1)--[out=280, in=260, min distance=0.6cm](x2),
				(x) --[out=135, in=180, min distance=0.1cm] (y) --[out=0, in=45, min distance=0.1cm](x),
			};
		\end{feynman}
	\end{tikzpicture}
	-\frac{\lambda^2}{4}\begin{tikzpicture}[baseline=(z)]
		\begin{feynman}[inline=(a)]
			\vertex (a);
			\vertex [right=1.3cm of a] (b);
			\tikzfeynmanset{every vertex=dot}
			\vertex [right=0.1cm of a, label=90:$x_1$] (x1);
			\vertex [right=1cm of x1, label=90:$x_2$] (x2);
			\vertex [below right=0.8cm and 0.5cm of x1] (x);
			\vertex [below right =0.5cm and 0.5cm of x1] (y);
			\tikzfeynmanset{every vertex={empty dot,minimum size=0mm}}
			\vertex [below right =0.2cm and 0.5cm of x1] (z);
			\diagram* {
				(a)--[double](x1)--[double](x2)--[double](b),
				(x1)--[out=280, in=260, min distance=1cm](x2),
				(x) --[out=135, in=225, min distance=0.1cm] (y)  --[out=315, in=45, min distance=0.1cm] (x),
				(y) --[out=135, in=180, min distance=0.1cm] (z) --[out=0,in=45, min distance=0.1cm] (y),
			};
		\end{feynman}
	\end{tikzpicture}\\
	&-\frac{\lambda^2}{4}\begin{tikzpicture}[baseline=(h)]
		\begin{feynman}[inline=(a)]
			\vertex (a);
			\vertex [right=1.3cm of a] (b);
			\tikzfeynmanset{every vertex=dot}
			\vertex [right=0.1cm of a, label=90:$x_1$] (x1);
			\vertex [right=1cm of x1, label=90:$x_2$] (x2);
			\vertex [below right=0.45cm and 0.3cm of x1] (x);
			\vertex [below right=0.45cm and 0.7cm of x1] (y);
			\tikzfeynmanset{every vertex={empty dot,minimum size=0mm}}
			\vertex [below right =0.2cm and 0.3cm of x1] (h);
			\vertex [below right =0.2cm and 0.7cm of x1] (g);
			\diagram* {
				(a)--[double](x1)--[double](x2)--[double](b),
				(x1)--[out=280, in=260, min distance=0.6cm](x2),
				(x) --[out=135, in=180, min distance=0.1cm] (h) --[out=0, in=45, min distance=0.1cm](x),
				(y) --[out=135, in=180, min distance=0.1cm] (g) --[out=0, in=45, min distance=0.1cm](y),
			};
		\end{feynman}
	\end{tikzpicture}-\frac{\lambda^2}{6}\begin{tikzpicture}[baseline=(h)]
		\begin{feynman}[inline=(a)]
			\vertex (a);
			\vertex [right=1.3cm of a] (b);
			\tikzfeynmanset{every vertex=dot}
			\vertex [right=0.1cm of a, label=90:$x_1$] (x1);
			\vertex [right=1cm of x1, label=90:$x_2$] (x2);
			\vertex [below right=0.45cm and 0.3cm of x1] (x);
			\vertex [below right=0.45cm and 0.7cm of x1] (y);
			\tikzfeynmanset{every vertex={empty dot,minimum size=0mm}}
			\vertex [below right =0.2cm and 0.5cm of x1] (h);
			\diagram* {
				(a)--[double](x1)--[double](x2)--[double](b),
				(x1)--[out=280, in=260, min distance=0.6cm](x2),
				(x)--[out=85, in=95, min distance=0.2cm](y)--[out=265, in=275, min distance=0.2cm](x),
			};
		\end{feynman}
	\end{tikzpicture}
\end{align}
\begin{align}
	\expect{\Op(\un{x}_1)\Op(\un{x}_2)\Op(\un{x}_3)\Op(\un{x}_4)}
	=&\frac{\delta^4\Psi[\phi_0]}{\delta\phi_0(\un{x}_1)\delta\phi_0(\un{x}_2)\delta\phi_0(\un{x}_3)\delta\phi_0(\un{x}_4)}\\
	=&3\times\begin{tikzpicture}[baseline=(base)]
		\begin{feynman}[inline=(base)]
			\vertex (a);
			\vertex [right=1.9cm of a] (b);
			\tikzfeynmanset{every vertex=dot}
			\vertex [right=0.1cm of a, label=90:$x_1$] (x1);
			\vertex [right=0.6cm of x1, label=90:$x_2$] (x2);
			\vertex [right=0.4cm of x2, label=90:$x_3$] (x3);
			\vertex [right=0.6cm of x3, label=90:$x_4$] (x4);
			\tikzfeynmanset{every vertex={empty dot,minimum size=0mm}}
			\vertex [below=0.1cm of x1] (base);
			\diagram* {
				(a)--[double](x1)--[double](x2)--[double](x3)--[double](x4)--[double](b),
				(x1)--[out=280, in=260, min distance=0.3cm](x2),
				(x3)--[out=280, in=260, min distance=0.3cm](x4),
			};
		\end{feynman}
	\end{tikzpicture}-i\lambda\left(3\times\begin{tikzpicture}[baseline=(base)]
		\begin{feynman}[inline=(base)]
			\vertex (a);
			\vertex [right=1.9cm of a] (b);
			\tikzfeynmanset{every vertex=dot}
			\vertex [right=0.1cm of a, label=90:$x_1$] (x1);
			\vertex [right=0.6cm of x1, label=90:$x_2$] (x2);
			\vertex [right=0.4cm of x2, label=90:$x_3$] (x3);
			\vertex [right=0.6cm of x3, label=90:$x_4$] (x4);
			\vertex [below right=0.25cm and 0.3cm of x1] (x);
			\tikzfeynmanset{every vertex={empty dot,minimum size=0mm}}
			\vertex [below right=0.1cm and 0.3cm of x1] (base);
			\vertex [below right=0.05cm and 0.3cm of x1] (y);
			\diagram* {
				(a)--[double](x1)--[double](x2)--[double](x3)--[double](x4)--[double](b),
				(x1)--[out=280, in=260, min distance=0.3cm](x2),
				(x3)--[out=280, in=260, min distance=0.3cm](x4),
				(x)--[out=120, in=180, min distance=0.1cm](y)--[out=0, in=60, min distance=0.1cm](x),
			};
		\end{feynman}
	\end{tikzpicture}+\begin{tikzpicture}[baseline=(base)]
		\begin{feynman}[inline=(a)]
			\vertex (a);
			\vertex [right=1.7cm of a] (b);
			\tikzfeynmanset{every vertex=dot}
			\vertex [right=0.1cm of a, label=90:$x_1$] (x1);
			\vertex [below right=0.4cm and 0.75cm of x1] (x);
			\vertex [right=0.5cm of x1, label=90:$x_2$] (x2);
			\vertex [right=0.5cm of x2, label=90:$x_3$] (x3);
			\vertex [right=0.5cm of x3, label=90:$x_4$] (x4);
			\tikzfeynmanset{every vertex={empty dot,minimum size=0mm}}
			\vertex [below=0.2cm of x1] (base);
			\diagram* {
				(a)--[double](x1)--[double](x2)--[double](x3)--[double](x4)--[double](b),
				(x1)--(x)--(x2),
				(x3)--(x)--(x4),
			};
		\end{feynman}
	\end{tikzpicture}\right)\\
	&-3\times\lambda^2\left(\frac12\begin{tikzpicture}[baseline=(y)]
		\begin{feynman}[inline=(y)]
			\vertex (a);
			\vertex [right=1.9cm of a] (b);
			\tikzfeynmanset{every vertex=dot}
			\vertex [right=0.1cm of a, label=90:$x_1$] (x1);
			\vertex [right=0.6cm of x1, label=90:$x_2$] (x2);
			\vertex [right=0.4cm of x2, label=90:$x_3$] (x3);
			\vertex [right=0.6cm of x3, label=90:$x_4$] (x4);
			\vertex [below right=0.5cm and 0.3cm of x1] (x);
			\vertex [below right=0.25cm and 0.3cm of x1] (y);
			\tikzfeynmanset{every vertex={empty dot,minimum size=0mm}}
			\vertex [below right=0.05cm and 0.3cm of x1] (z);
			\diagram* {
				(a)--[double](x1)--[double](x2)--[double](x3)--[double](x4)--[double](b),
				(x1)--[out=280, in=260, min distance=0.6cm](x2),
				(x3)--[out=280, in=260, min distance=0.6cm](x4),
				(x) --[out=135, in=225, min distance=0.1cm] (y)  --[out=315, in=45, min distance=0.1cm] (x),
				(y) --[out=135, in=180, min distance=0.1cm] (z) --[out=0,in=45, min distance=0.1cm] (y),
			};
		\end{feynman}
	\end{tikzpicture}+\frac12\begin{tikzpicture}[baseline=(base)]
		\begin{feynman}[inline=(base)]
			\vertex (a);
			\vertex [right=1.9cm of a] (b);
			\tikzfeynmanset{every vertex=dot}
			\vertex [right=0.1cm of a, label=90:$x_1$] (x1);
			\vertex [right=0.6cm of x1, label=90:$x_2$] (x2);
			\vertex [right=0.4cm of x2, label=90:$x_3$] (x3);
			\vertex [right=0.6cm of x3, label=90:$x_4$] (x4);
			\vertex [below right=0.25cm and 0.2cm of x1] (x);
			\vertex [below right=0.25cm and 0.4cm of x1] (z);
			\tikzfeynmanset{every vertex={empty dot,minimum size=0mm}}
			\vertex [below right=0.25cm and 0.3cm of x1] (base);
			\vertex [below right=0.05cm and 0.2cm of x1] (y);
			\vertex [below right=0.05cm and 0.4cm of x1] (y2);
			\diagram* {
				(a)--[double](x1)--[double](x2)--[double](x3)--[double](x4)--[double](b),
				(x1)--[out=280, in=260, min distance=0.3cm](x2),
				(x3)--[out=280, in=260, min distance=0.3cm](x4),
				(x)--[out=120, in=180, min distance=0.05cm](y)--[out=0, in=60, min distance=0.05cm](x),
				(z)--[out=120, in=180, min distance=0.05cm](y2)--[out=0, in=60, min distance=0.05cm](z),
			};
		\end{feynman}
	\end{tikzpicture}+\frac13\begin{tikzpicture}[baseline=(base)]
		\begin{feynman}[inline=(base)]
			\vertex (a);
			\vertex [right=1.9cm of a] (b);
			\tikzfeynmanset{every vertex=dot}
			\vertex [right=0.1cm of a, label=90:$x_1$] (x1);
			\vertex [right=0.6cm of x1, label=90:$x_2$] (x2);
			\vertex [right=0.4cm of x2, label=90:$x_3$] (x3);
			\vertex [right=0.6cm of x3, label=90:$x_4$] (x4);
			\vertex [below right=0.25cm and 0.15cm of x1] (x);
			\vertex [below right=0.25cm and 0.45cm of x1] (z);
			\tikzfeynmanset{every vertex={empty dot,minimum size=0mm}}
			\vertex [below right=0.25cm and 0.3cm of x1] (base);
			\diagram* {
				(a)--[double](x1)--[double](x2)--[double](x3)--[double](x4)--[double](b),
				(x1)--[out=280, in=260, min distance=0.3cm](x2),
				(x3)--[out=280, in=260, min distance=0.3cm](x4),
				(x)--[out=85, in=95, min distance=0.15cm](z)--[out=265, in=275, min distance=0.15cm](x),
			};
		\end{feynman}
	\end{tikzpicture}\right.\\
	&\left.+\frac14\begin{tikzpicture}[baseline=(base)]
		\begin{feynman}[inline=(base)]
			\vertex (a);
			\vertex [right=1.9cm of a] (b);
			\tikzfeynmanset{every vertex=dot}
			\vertex [right=0.1cm of a, label=90:$x_1$] (x1);
			\vertex [right=0.6cm of x1, label=90:$x_2$] (x2);
			\vertex [right=0.4cm of x2, label=90:$x_3$] (x3);
			\vertex [right=0.6cm of x3, label=90:$x_4$] (x4);
			\vertex [below right=0.25cm and 0.3cm of x1] (x);
			\vertex [below right=0.25cm and 0.3cm of x3] (z);
			\tikzfeynmanset{every vertex={empty dot,minimum size=0mm}}
			\vertex [below right=0.1cm and 0.3cm of x1] (base);
			\vertex [below right=0.05cm and 0.3cm of x1] (y);
			\vertex [below right=0.05cm and 0.3cm of x3] (w);
			\diagram* {
				(a)--[double](x1)--[double](x2)--[double](x3)--[double](x4)--[double](b),
				(x1)--[out=280, in=260, min distance=0.3cm](x2),
				(x3)--[out=280, in=260, min distance=0.3cm](x4),
				(x)--[out=120, in=180, min distance=0.1cm](y)--[out=0, in=60, min distance=0.1cm](x),
				(z)--[out=120, in=180, min distance=0.1cm](w)--[out=0, in=60, min distance=0.1cm](z),
			};
		\end{feynman}
	\end{tikzpicture}+4\times\frac12\begin{tikzpicture}[baseline=(base)]
		\begin{feynman}[inline=(a)]
			\vertex (a);
			\vertex [right=1.7cm of a] (b);
			\tikzfeynmanset{every vertex=dot}
			\vertex [right=0.1cm of a, label=90:$x_1$] (x1);
			\vertex [below right=0.4cm and 0.75cm of x1] (x);
			\vertex [right=0.5cm of x1, label=90:$x_2$] (x2);
			\vertex [right=0.5cm of x2, label=90:$x_3$] (x3);
			\vertex [right=0.5cm of x3, label=90:$x_4$] (x4);
			\vertex [below right=0.2cm and 0.375cm of x1] (z);
			\tikzfeynmanset{every vertex={empty dot,minimum size=0mm}}
			\vertex [below=0.2cm of x1] (base);
			\vertex [below right=0.4cm and 0.25cm of x1] (w);
			\diagram* {
				(a)--[double](x1)--[double](x2)--[double](x3)--[double](x4)--[double](b),
				(x1)--(x)--(x2),
				(x3)--(x)--(x4),
				(z)--[out=170, in=150, min distance=0.1cm](w)--[out=315, in=300, min distance=0.1cm](z),
			};
		\end{feynman}
	\end{tikzpicture}+3\times\frac12\begin{tikzpicture}[baseline=(base)]
		\begin{feynman}[inline=(base)]
			\vertex (a);
			\vertex [right=1.9cm of a] (b);
			\tikzfeynmanset{every vertex=dot}
			\vertex [right=0.1cm of a, label=90:$x_1$] (x1);
			\vertex [right=0.4cm of x1, label=90:$x_2$] (x2);
			\vertex [right=0.8cm of x2, label=90:$x_3$] (x3);
			\vertex [right=0.4cm of x3, label=90:$x_4$] (x4);
			\vertex [below right=0.4cm and 0.5cm of x1] (x);
			\vertex [below left=0.4cm and 0.5cm of x4] (z);
			\tikzfeynmanset{every vertex={empty dot,minimum size=0mm}}
			\vertex [below=0.2cm of x1] (base);
			\diagram* {
				(a)--[double](x1)--[double](x2)--[double](x3)--[double](x4)--[double](b),
				(x1)--[out=270, in=170, min distance=0.1cm](x)--[out=120, in=270, min distance=0.1cm](x2),
				(x3)--[out=270, in=60, min distance=0.1cm](z)--[out=10, in=270, min distance=0.1cm](x4),
				(x)--[out=80, in=100, min distance=0.3cm](z)--[out=260, in=280, min distance=0.3cm](x),
			};
		\end{feynman}
	\end{tikzpicture}\right)
\end{align}
\subsubsection{Two point function}
Now we have all the parts in place to calculate the two point function of the dual conformal field theory by explicitly integrating over the vertices in the diagrams. We will perform all the calculations directly in de Sitter space however we will find out that all the results are exactly what you would get from taking the results from \cite{Bertan:2018khc} and do the analytic continuation. 

\paragraph{Tadpole}
The second diagram in the two point function is the tadpole and is given by the following integral:
\begin{align}
	\begin{tikzpicture}[baseline=(y)]
		\begin{feynman}[inline=(a)]
			\vertex (a);
			\vertex [right=1.3cm of a] (b);
			\tikzfeynmanset{every vertex=dot}
			\vertex [right=0.1cm of a, label=90:$x_1$] (x1);
			\vertex [right=1cm of x1, label=90:$x_2$] (x2);
			\vertex [below right=0.5cm and 0.5cm of x1] (x);
			\tikzfeynmanset{every vertex={empty dot,minimum size=0mm}}
			\vertex [below right =0.2cm and 0.5cm of x1] (y);
			\diagram* {
				(a)--[double](x1)--[double](x2)--[double](b),
				(x1)--[out=280, in=260, min distance=0.6cm](x2),
				(x) --[out=135, in=180, min distance=0.1cm] (y) --[out=0, in=45, min distance=0.1cm](x),
			};
		\end{feynman}
	\end{tikzpicture}=&\int\dd^4x\sqrt{g(x)}K(\un{x}_1,x)iG(x,x)K(x,\un{x}_2)=iG(x,x)M(\un{x}_1,\un{x}_2)\label{eq:tadpole}
\end{align}
First we calculate the massshift $M(\un{x}_1,\un{x}_2)$ which is given by the integral:
\begin{align}
	M(\un{x}_1,\un{x}_2)&=-\frac{\ell^4}{\pi^4\epsilon^2}\int_{-\infty}^{0}\frac{\dd\eta}{\eta^4}\int\dd^3 \un{x}
	\frac{\eta^4}{(\eta^2-(\un{x}_1-\un{x})^2-i\epsilon)^2(\eta^2-(\un{x}_2-\un{x})^2-i\epsilon)^2}\label{eq:massshift1}
\end{align}
To simplify the calculation we use translation invariance to shift $\un{x}_1$ and $\un{x}_2$ by $\un{x}_2$ and we get $\un{x}'_1=\un{x}_1-\un{x}_2$ and $\un{x}'_2=0$. Now we can go to spherical coordinates with $r=\abs{\un{x}}$ and the determinant becomes  $\sqrt{g(x)}=\frac{\ell^4}{\eta^4}r^2\sin\theta$:
\begin{align}
	M(\un{x}_1,\un{x}_2)&=-\frac{\ell^4}{\pi^4\epsilon^2}\int_{-\infty}^{0}\dd\eta\int_0^{\pi}\dd\theta\int_0^{\infty}\dd r
	\frac{2\pi r^2\sin\theta}{(\eta^2-\abs{\un{x}'_1}^2+2\abs{\un{x}'_1}r\cos\theta-r^2-i\epsilon)^2(\eta^2-r^2-i\epsilon)^2}\\
	&=-\frac{2\ell^4}{\pi^3\epsilon^2}\int_{-\infty}^{0}\dd\eta\int_{-1}^{+1}\dd u\int_0^{\infty}\dd r
	\frac{2\pi r^2}{(\eta^2-\abs{\un{x}'_1}^2+2\abs{\un{x}'_1}r u-r^2-i\epsilon)^2(\eta^2-r^2-i\epsilon)^2}\\
	&=-\frac{4\ell^4}{\pi^3\epsilon^2}\int\limits_{-\infty}^{0}\dd\eta\int\limits_0^{\infty}\dd r
	\frac{r^2}{(\eta^2-(\abs{\un{x}'_1}+r)^2-i\epsilon)(\eta^2-(\abs{\un{x}'_1}-r)^2-i\epsilon)(\eta^2-r^2-i\epsilon)^2}
\end{align}
It is clear that the argument of the integral is invariant under $r\rightarrow -r$ so the integration can be done over the whole real axis.
By continuing the domain of $r$ to the complex plane we see that the integrand of $M$ has the following poles in $r$:
\begin{align}
	M(\un{x}_1,\un{x}_2)=&-\frac{4\ell^4}{\pi^3\epsilon^2}\int\limits_{-\infty}^{0}\dd\eta\int\limits_{-\infty}^{\infty}\dd r r^2\frac{1}{2}
	\frac{(\eta-i\varepsilon-\abs{\un{x}'_1}-r)^{-1}(\eta-i\varepsilon+\abs{\un{x}'_1}-r)^{-1}(\eta-i\varepsilon-r)^{-2}}{(\eta-i\varepsilon+\abs{\un{x}'_1}+r)(\eta-i\varepsilon-\abs{\un{x}'_1}+r)(\eta-i\varepsilon+r)^2}\\
	r_1=&-\eta-x_1+i\varepsilon\\
	r_2=&-\eta+x_1+i\varepsilon\\
	r_3=&-\eta+i\varepsilon\\
	r_4=&\eta-x_1-i\varepsilon\\
	r_5=&\eta+x_1-i\varepsilon\\
	r_6=&\eta-i\varepsilon
\end{align}
By either closing the integration contour in the upper or lower half we see that we include either the poles $r_1,r_2,r_3$ or $r_4,r_5,r_6$ as the sign of $i\varepsilon$ is fixed by the sign convention of the metric. The choice does not matter so we pick the last three poles and integrate along the following contour:
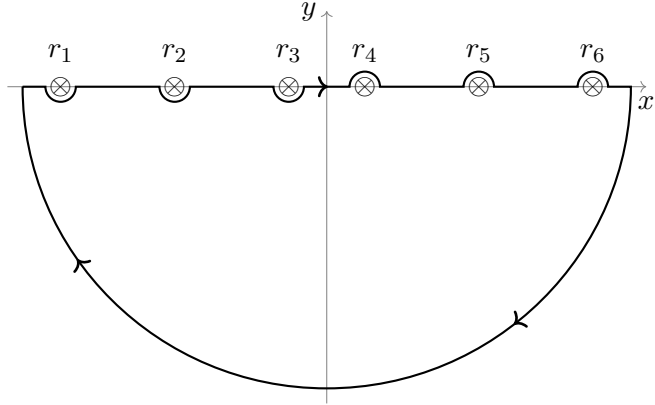
\begin{center}
	\begin{tikzpicture}[decoration={markings,
			mark=at position 4.7cm with {\arrow[line width=1pt]{>}},
			mark=at position 13cm with {\arrow[line width=1pt]{>}},
			mark=at position 19.5cm with {\arrow[line width=1pt]{>}},
		}
		]
		\draw[help lines,->] (-4.2,0) -- (4.2,0) coordinate (xaxis);
		\draw[help lines,->] (0,-4.2) -- (0,1) coordinate (yaxis);
		
		\path[draw,line width=0.8pt,postaction=decorate] (-4,0)--(-3.7,0) arc (180:360:0.2)--(-2.2,0) arc (180:360:0.2)--(-0.7,0) arc (180:360:0.2)
		--(0.3,0) arc (180:0:0.2)--(1.8,0) arc (180:0:0.2)--(3.3,0) arc (180:0:0.2)--(4,0) arc (360:180:4);
		
		\node[below] at (xaxis) {$x$};
		\node[left] at (yaxis) {$y$};
		\node at (-3.5,0) {$\otimes$};
		\node at (-2,0) {$\otimes$};
		\node at (-0.5,0) {$\otimes$};
		\node at (0.5,0) {$\otimes$};
		\node at (2,0) {$\otimes$};
		\node at (3.5,0) {$\otimes$};
		\node[above=0.2] at (-3.5,0) {$r_1$};
		\node[above=0.2] at (-2,0) {$r_2$};
		\node[above=0.2] at (-0.5,0) {$r_3$};
		\node[above=0.2] at (0.5,0) {$r_4$};
		\node[above=0.2] at (2,0) {$r_5$};
		\node[above=0.2] at (3.5,0) {$r_6$};
	\end{tikzpicture}
	\label{fig:contour}
\end{center}
We use the Residue theorem to solve the $r$ integral:
\begin{align}
	M(\un{x}_1,\un{x}_2)=&-\frac{4\ell^4}{\pi^3\epsilon^2}\int\limits_{-\infty}^{\epsilon}\dd\eta(-2\pi i)\left(\frac{(\abs{\un{x}'_1}-\eta )}{16\eta  \abs{\un{x}'_1}^3(\abs{\un{x}'_1}-2 \eta )^2}+\frac{(\eta +\abs{\un{x}'_1})}{16 \eta  \abs{\un{x}'_1}^3(2 \eta +\abs{\un{x}'_1})^2}\right.\\
	&+\left.\frac{1}{8 \eta  (\abs{\un{x}'_1}-2 \eta )^2 (2 \eta +\abs{\un{x}'_1})^2}\right)\\
	=&\frac{2 i \ell ^4}{\pi ^2\epsilon ^2}\int_{-\infty}^{\epsilon}\dd\eta\frac{1}{\eta\left(\abs{\un{x}'_1}^2-4 \eta ^2\right)^2}
	=\frac{i\ell^4}{\pi^2\epsilon^2\abs{\un{x}'_1}^4}\left(1+\ln\left(-\frac{4\epsilon^2}{\abs{\un{x}'_1}^2}\right)\right)
	+\Op(\epsilon)\\
	=&\frac{i\ell^4}{\pi^2\epsilon^2\abs{\un{x}_1-\un{x}_2}^4}\left(1+\ln\left(-\frac{4\epsilon^2}{\abs{\un{x}_1-\un{x}_2}^2}\right)\right)
	+\Op(\epsilon)
\end{align}
$\epsilon$ in this case is the cutoff before $\eta\rightarrow 0$ and the expansion is done around $\epsilon=0$.

Now that we calculated the massshift we can take a closer look at the first part of \eqref{eq:tadpole} which is the propagator at coinciding points.
In case of colliding points the geodesic distance goes to 1:
\begin{align}
	P_{xy}=\frac{2\eta_x\eta_y}{2\eta_x\eta_y+(\eta_x-\eta_y)^2-(x-y)^2}\stackrel{(\eta_x,x)\rightarrow (\eta_y,y)}{\longrightarrow} 1
\end{align}
Therefore the propagator diverges at coinciding points. To regularize this divergence we do the following replacement with small $\delta$ (as it was done in \cite{Bertan:2018afl}):
\begin{align}
	P&\rightarrow \frac{P}{1+\delta}\\
	\Rightarrow iG(x,y)&\rightarrow-\frac{1}{4\pi^2\ell^2}\frac{P^2}{(1+\delta+P)(1+\delta-P)}\stackrel{P\rightarrow 1}{\longrightarrow}-\frac{1}{4\pi^2\ell^2\delta(2+\delta)}
\end{align}
This regularization corresponds to carving out a ball of radius $\delta$ around the coinciding point and rescaling everything by $(1+\delta)^{-1}$. The complete one-loop tadpole diagram is then given by:
\begin{align}
	T_1(\un{x},\un{y})=\begin{tikzpicture}[baseline=(y)]
		\begin{feynman}[inline=(a)]
			\vertex (a);
			\vertex [right=1.3cm of a] (b);
			\tikzfeynmanset{every vertex=dot}
			\vertex [right=0.1cm of a, label=90:$x_1$] (x1);
			\vertex [right=1cm of x1, label=90:$x_2$] (x2);
			\vertex [below right=0.5cm and 0.5cm of x1] (x);
			\tikzfeynmanset{every vertex={empty dot,minimum size=0mm}}
			\vertex [below right =0.2cm and 0.5cm of x1] (y);
			\diagram* {
				(a)--[double](x1)--[double](x2)--[double](b),
				(x1)--[out=280, in=260, min distance=0.6cm](x2),
				(x) --[out=135, in=180, min distance=0.1cm] (y) --[out=0, in=45, min distance=0.1cm](x),
			};
		\end{feynman}
	\end{tikzpicture}=&iG(x,x)\frac{M(\un{x},\un{y})}{(1+\delta)^4}=
	-\frac{i\ell^2}{4\pi^4\epsilon^2(1+\delta)^4\delta(2+\delta)}
	\frac{1}{\abs{\un{x}-\un{y}}^4}\left(1+\ln\left(-\frac{4\epsilon^2}{\abs{\un{x}-\un{y}}^2}\right)\right)\\
	\stackrel{\delta\rightarrow 0}{\longrightarrow}&i\left(\frac{1}{\delta}-\frac92\right)\frac{\ell^2}{8\pi^4\epsilon^2}
	\frac{1}{\abs{\un{x}-\un{y}}^4}\left(1+\ln\left(-\frac{4\epsilon^2}{\abs{\un{x}-\un{y}}^2}\right)\right)
\end{align}

\paragraph{Double tadpole}
The next term in the expansion of the two point function is the double tadpole which is given by:
\begin{align}
	T_2(\un{x}_1,\un{x}_2)=\begin{tikzpicture}[baseline=(z)]
		\begin{feynman}[inline=(a)]
			\vertex (a);
			\vertex [right=1.3cm of a] (b);
			\tikzfeynmanset{every vertex=dot}
			\vertex [right=0.1cm of a, label=90:$x_1$] (x1);
			\vertex [right=1cm of x1, label=90:$x_2$] (x2);
			\vertex [below right=0.8cm and 0.5cm of x1] (x);
			\vertex [below right =0.5cm and 0.5cm of x1] (y);
			\tikzfeynmanset{every vertex={empty dot,minimum size=0mm}}
			\vertex [below right =0.2cm and 0.5cm of x1] (z);
			\diagram* {
				(a)--[double](x1)--[double](x2)--[double](b),
				(x1)--[out=280, in=260, min distance=1cm](x2),
				(x) --[out=135, in=225, min distance=0.1cm] (y)  --[out=315, in=45, min distance=0.1cm] (x),
				(y) --[out=135, in=180, min distance=0.1cm] (z) --[out=0,in=45, min distance=0.1cm] (y),
			};
		\end{feynman}
	\end{tikzpicture}&=\int\dd^4x\sqrt{g(x)}\int\dd^4y\sqrt{g(y)}K(\un{x}_1,x)i^2G^2(x,y)iG(y,y)K(x,\un{x}_2)
\end{align}
Again we use the regularization $P\rightarrow P/(1+\delta)$. Then the different parts in the integral become:
\begin{align}
	iG(y,y)&\rightarrow-\frac{1}{4\pi^2\ell^2\delta(2+\delta)}\\
	i^2G^2(x,y)&\rightarrow\frac{1}{(4\pi^2\ell^2)^2}\frac{P^4_{xy}}{((1+\delta)^2-P_{xy}^2)^2}\\
	K(\un{x}_1,x)K(x,\un{x}_2)&\rightarrow\frac{1}{(1+\delta)^4}K(\un{x}_1,x)K(x,\un{x}_2)
\end{align}
The whole integral becomes:
\begin{align}
	T_2(x_1,x_2)=&-\frac{1}{(4\pi^2\ell^2)^3}
	\frac{1}{\delta(2+\delta)(1+\delta)^4}
	\int\dd^4 x\sqrt{g(x)}K(\un{x}_1,x)K(x,\un{x}_2)\\
	&\times\underbrace{\int\dd^4 y\sqrt{g(y)}\frac{P^4_{xy}}{(1+\delta-P_{xy})^2(1+\delta+P_{xy})^2}}_{:=\mathcal{K}}
\end{align}
We concentrate now on the last part of the integral. In local coordinates this is given by:
\begin{align}
	\mathcal{K}&=\ell^4\int\dd^4 y\sqrt{g(y)}\frac{P^4_{xy}}{(1+\delta-P_{xy})^2(1+\delta+P_{xy})^2}\\
	&=\int\dd^4 y\frac{16 \eta_x^4}{\left((1+\delta)^2(\eta_x^2+\eta_y^2-(x-y)^2)^2-4\eta_x^2\eta_y^2\right)^2}
\end{align}
To show that this part is independent of $(\eta_x,x)$ we use translation invariance to shift the spatial part of the  integral to $(\eta_y,y')=(\eta_y,y+x)$. Then $\mathcal{K}$ becomes:
\begin{align}
	\mathcal{K}&=\ell^4\int\dd^4 y'\frac{16 \eta_x^4}{\left((1+\delta)^2(\eta_x^2+\eta_y^2-y'^2)^2-4\eta_x^2\eta_y^2\right)^2}\\	&=\ell^4\int\dd^4y\frac{16\eta_x^4}{\left[((1+\delta)(\eta_x^2+\eta_y^2-y'^2)+2\eta_x\eta_y)((1+\delta)(\eta_x^2+\eta_y^2-y'^2)-2\eta_x\eta_y)\right]}
\end{align}
To see that this is also independent of $\eta_x$ we use scale invariance. It is easy to see that any rescaling of $\eta_x\rightarrow\lambda\eta_x$ can be undone be rescaling 
$(\eta_y,y')\rightarrow(\lambda\eta_y,\lambda y')$. Therefore we can fix $\eta_x$ to any random value so we choose $\eta_x=-1$. Therefore we get:
\begin{align}
	\mathcal{K}&=\ell^4\int\dd^4y\frac{16}{\left[((1+\delta)(1+\eta_y^2-y'^2)+2\eta_y)((1+\delta)(1+\eta_y^2-y'^2)-2\eta_y)\right]}
\end{align}
So $T_2$ factorizes into:
\begin{align}
	T_2(\un{x}_1,\un{x}_2)&=\frac{1}{(4\pi^2\ell^2)^3}\frac{1}{\delta(2+\delta)(1+\delta)^4}\times M(\un{x}_1,\un{x}_2)\times\mathcal{K}
\end{align}
As the integral $\mathcal{K}$ is symmetric under $\eta_y\rightarrow-\eta_y$ we can write it as an integral over $\eta_y\in]-\infty,\infty[$. We can do the spatial part of the integral in spherical coordinates. After integrating over the angular part we get:
\begin{align}
	\mathcal{K}&=32\pi\ell^4\int_{-\infty}^{\infty}\dd\eta\int_{0}^{\infty}\dd r\frac{r^2}{\left[((1+\delta)(\eta^2+1-r^2)+2\eta)((1+\delta)(\eta^2+1-r^2)-2\eta)\right]^2}
\end{align}
To execute the $r$ integral we use again the residue theorem. After that the $\eta$ integral can be calculated straightforwardly:
\begin{align}
	\mathcal{K}&=32\pi^2\ell^4 i\int_{-\infty}^{\infty}\dd\eta \frac{1}{64(1+\delta)^{3/2}\eta^3}\left(\frac{(1+\delta)(1+\eta^2)-\eta}{\sqrt{(\eta-1)^2+\delta(1+\eta^2)}}
	-\frac{(1+\delta)(1+\eta^2)+\eta}{\sqrt{(\eta+1)^2+\delta(1+\eta^2)}}\right)\\
	&=\frac{i\pi^2\ell^4}{2(1+\delta)^{3/2}}\int_{-\infty}^{\infty}\dd\eta \frac{1}{\eta^3}\left(\frac{(1+\delta)(1+\eta^2)-\eta}{\sqrt{(\eta-1)^2+\delta(1+\eta^2)}}
	-\frac{(1+\delta)(1+\eta^2)+\eta}{\sqrt{(\eta+1)^2+\delta(1+\eta^2)}}\right)\\
	&=-i\pi^2\ell^4\frac{2(1+\delta)+(2+\delta(2+\delta))\ln\left(\frac{\delta}{2+\delta}\right)}{2(1+\delta)^3}
\end{align}
Therefore the complete double tadpole integral becomes:
\begin{align}
	T_2(\un{x}_1,\un{x}_2)&=-\frac{i\pi^2\ell^4}{(4\pi^2\ell^2)^3}
	\frac{2(1+\delta)+(2+\delta(2+\delta))\ln\left(\frac{\delta}{2+\delta}\right)}{2\delta(2+\delta)(1+\delta)^7}\times M(\un{x}_1,\un{x}_2)\\
\end{align}
For small $\delta$ this becomes:
\begin{align}
	T_2(\un{x}_1,\un{x}_2)&=\frac{i\pi^2}{2(4\pi^2)^3\ell^2}\left(\frac{14+13\ln\frac{\delta}{2}}{2}-\frac{1+\ln\frac{\delta}{2}}{\delta}\right)\times M(\un{x}_1,\un{x}_2)+\Op(\delta)\\
	&=-\frac{\ell^2}{2(4\pi^2)^3\epsilon^2}\left(\frac{14+13\ln\frac{\delta}{2}}{2}-\frac{1+\ln\frac{\delta}{2}}{\delta}\right)
	\frac{1}{\abs{\un{x}-\un{y}}^4}\left(1+\ln\left(-\frac{4\epsilon^2}{\abs{\un{x}-\un{y}}^2}\right)\right)
\end{align}

\paragraph{Sunrise diagram}
The next second order diagram is the so called sunrise diagram. It is given by the following integral:
\begin{align}
	S(\un{x}_1,\un{x}_2)=\begin{tikzpicture}[baseline=(h)]
		\begin{feynman}[inline=(a)]
			\vertex (a);
			\vertex [right=1.3cm of a] (b);
			\tikzfeynmanset{every vertex=dot}
			\vertex [right=0.1cm of a, label=90:$x_1$] (x1);
			\vertex [right=1cm of x1, label=90:$x_2$] (x2);
			\vertex [below right=0.45cm and 0.3cm of x1] (x);
			\vertex [below right=0.45cm and 0.7cm of x1] (y);
			\tikzfeynmanset{every vertex={empty dot,minimum size=0mm}}
			\vertex [below right =0.2cm and 0.5cm of x1] (h);
			\diagram* {
				(a)--[double](x1)--[double](x2)--[double](b),
				(x1)--[out=280, in=260, min distance=0.6cm](x2),
				(x)--[out=85, in=95, min distance=0.2cm](y)--[out=265, in=275, min distance=0.2cm](x),
			};
		\end{feynman}
	\end{tikzpicture}&=\int\dd^4x\int\dd^4y\sqrt{g(x)g(y)}K(\un{x}_1,x)i^3G^3(x,y)K(y,\un{x}_2)
\end{align}
Again using the same regularization as above this integral becomes:
\begin{align}
	S(\un{x}_1,\un{x}_2)=&\frac{1}{(4\pi^2\ell^2)^3(1+\delta)^4}\frac{1}{\pi^4\epsilon^2}\int\dd^4x\sqrt{g(x)}\bar{P}^2_{\un{x}_1 y}
	\underbrace{\int\dd^4y\sqrt{g(y)}\frac{P^6_{xy}}{(1+\delta+P_{xy})^3(1+\delta-P_{xy})^3}\bar{P}^2_{\un{x}_2 x}}_{:=J}
\end{align}
We can split off one leg from the integral and calculate $J$ first and then attaching the missing leg in the final step:
\begin{align}
	J(y,\un{x}_2)&=\frac{1}{(4\pi^2\ell^2)^3(1+\delta)^4}\frac{1}{\pi^4\epsilon^2}\int\dd^4y\sqrt{g(y)}\frac{P^6_{xy}}{(1+\delta+P_{xy})^3(1+\delta-P_{xy})^3}\bar{P}^2_{y,\un{x}_2'}\\
\end{align}
We shift $\un{x}_2$ and $\un{x}$ by $-\un{x}_2$ to $\un{x}_2'=\un{x}_2-\un{x}_2=0$ and $\un{x}'=\un{x}-\un{x}_2$. The integral then becomes:
\begin{align}
	J=\frac{1}{(4\pi^2\ell^2)^3(1+\delta)^4}\frac{1}{\pi^4\epsilon^2}\int\dd^4y\sqrt{g(y)}\frac{P^6_{x'y}}{(1+\delta+P_{x'y})^3(1+\delta-P_{x'y})^3}\bar{P}^2_{y \un{x}_2'}
\end{align}
Using inversion invariance to simplifies the integral even further. Inversion in de Sitter space is given by:
\begin{align}
	&\eta\rightarrow\frac{\eta'}{\eta'^2-\un{x}'^2};\qquad x_i\rightarrow\frac{x'_i}{\eta'^2-\un{x}'^2}\\
	&\Rightarrow P_{xy}=\frac{2\eta_x\eta_y}{\eta_x^2+\eta_y^2-(\un{x}-\un{y})^2}\rightarrow
	\frac{2\eta'_x\eta'_y}{{\eta'_x}^2+{\eta'_y}^2-(\un{x}'-\un{y}')^2}=P_{x'y'}\\
	&\Rightarrow\bar{P}_{y\un{x}_2}=\frac{\eta_y}{\eta_y^2-(\un{y}-\un{x}_2)^2}\rightarrow\frac{\eta'}{{\eta'_y}^2-\un{y}'^2}
	=\abs{\un{x}'_2}^2\bar{P}_{y'\un{x}'_2}
\end{align}
As the propagator in de Sitter space time is invariant under inversion we can use these identities to further simplify our integral for $J$. By shifting $\un{x}$ and $\un{x}_2$ by $-\un{x}_2$ we set $\un{x}_2'=0$. Now applying inversion of every point we send $\un{x}'_2\rightarrow\un{x}''_2=\frac{\un{x}'_2}{\abs{\un{x}'_2}^2}=\infty$. Then the bulk-to-boundary propagator becomes:
\begin{align}
	\bar{P}_{y\un{x}'_2}\rightarrow\abs{\un{x}''_2}^2\bar{P}_{y'\un{x}''_2}\stackrel{\abs{\un{x}''_2}\rightarrow\infty}{\longrightarrow}\eta'_y
\end{align}
After shifting $\un{y}''=\un{y}'+\un{x}''$ we can write $J$ as:
\begin{align}
	J=&\frac{2^6{\eta_x''}^6}{(4\pi^2)^3\ell^2(1+\delta)^4}\frac{1}{\pi^4\epsilon^2}\int\frac{\dd^4y'}{{\eta_y'}^4}
	\frac{{\eta_y'}^8((1+\delta)({\eta_x''}^2+{\eta_y'}^2-\abs{\un{y}'}^2)+2\eta_x''\eta_y')^{-3}}{((1+\delta)({\eta_x''}^2+{\eta_y'}^2-\abs{\un{y}'}^2)-2\eta_x''\eta_y')^3}\\
	=&\frac{4{\eta_x''}^6}{\pi^5\ell^2(1+\delta)^4}\frac{1}{\pi^4\epsilon^2}\int\limits_{-\infty}^{0}\dd\eta_y'\int\limits_0^{\infty}\dd r
	\frac{r^2 {\eta_y'}^4((1+\delta)({\eta_x''}^2+{\eta_y'}^2-r^2)+2\eta_x''\eta_y')^{-3}}{((1+\delta)({\eta_x''}^2+{\eta_y'}^2-r^2)-2\eta_x''\eta_y')^3}\\
	=&\frac{2{\eta_x''}^6}{\pi^5\ell^2(1+\delta)^4}\frac{1}{\pi^4\epsilon^2}\int\limits_{-\infty}^{+\infty}\dd\eta_y'\int\limits_0^{+\infty}\dd r
	\frac{r^2 {\eta_y'}^4((1+\delta)({\eta_x''}^2+{\eta_y'}^2-r^2)+2\eta_x''\eta_y')^{-3}}{((1+\delta)({\eta_x''}^2+{\eta_y'}^2-r^2)-2\eta_x''\eta_y')^3}
\end{align}
In the last step the fact was used that the integral is symmetric under $\eta_y\rightarrow-\eta_y$ to extend the limits of the integral.

The integrand has four poles of third order:
\begin{align}
	r_{\pm 1}&=\pm\sqrt{\frac{(1+\delta)({\eta_x''}^2+\eta_y^2)+2\eta_x''\eta_y}{1+\delta}}\\
	r_{\pm 2}&=\pm\sqrt{\frac{(1+\delta)({\eta_x''}^2+\eta_y^2)-2\eta_x''\eta_y}{1+\delta}}
\end{align}
We extend the integral to the complex plane and choose an integral contour such that the two poles with positive sign are included. Then performing the $\eta_y'$ integral gives:
\begin{align}
	J=&\frac{4i{\eta_x''}^6}{\pi^8\ell^2\epsilon^2(1+\delta)^4}\frac{6 \delta(\delta +2)+3\delta(\delta +1) (\delta +2) (\log (\delta )-\log (\delta +2))+2}{512{\eta_x''}^4 \delta  (\delta +1)^2 (\delta +2)}\\
	=&\frac{4i{\eta_x''}^2}{\pi^8\ell^2\epsilon^2}\frac{6\delta(\delta +2)+3\delta(\delta +1)(\delta +2)\ln\frac{\delta}{\delta +2}+2}{512\delta(\delta+1)^6(\delta+2)}
\end{align}
For the first order in $\delta$ this is:
\begin{align}
	J&=\frac{4i{\eta_x''}^2}{\pi^8\ell^2\epsilon^2}\frac{1}{2*4^4}\left(\frac{1}{\delta}+\frac{6\ln\frac{\delta}{2}-1}{2}\right)
	=\frac{2i{\eta_x''}^2}{(4\pi^2)^4\ell^2\epsilon^2}\left(\frac{1}{\delta}+\frac{6\ln\frac{\delta}{2}-1}{2}\right)
\end{align}
Now we can revert the inversion we did before the integration and take back the shift by $-\un{x}_2$:
\begin{align}
	\eta_x''&\rightarrow \frac{\eta_x}{\eta_x^2-\un{x}'^2}=\frac{\eta_x}{\eta_x^2-(\un{x}-\un{x}_2)^2}=\bar{P}_{x\un{x}_2}\\
	\Rightarrow J&=\frac{2i}{(4\pi^2)^4\ell^2\epsilon^2}\left(\frac{1}{\delta}+\frac{6\ln\frac{\delta}{2}-1}{2}\right)\bar{P}_{x\un{x}_2}^2
\end{align}
So finally by attaching the missing leg to $J$ we get the complete sunrise diagram:
\begin{align}
	S(\un{x}_1,\un{x}_2)&=\int\dd^4x\sqrt{g(x)}\bar{P}_{\un{x}_1 x}^2 J(x,\un{x}_2)=
	\frac{2i}{(4\pi^2)^4\ell^2\epsilon^2}\left(\frac{1}{\delta}+\frac{6\ln\frac{\delta}{2}-1}{2}\right)\int\dd^4x\sqrt{g(x)}\bar{P}_{\un{x}_1 x}^2\bar{P}_{x \un{x}_2}^2\\
	&=\frac{i\pi^2}{2(4\pi^2)^3\ell^2}\left(\frac{1}{\delta}+\frac{6\ln\frac{\delta}{2}-1}{2}\right)M(\un{x}_1,\un{x}_2)\\
	&=\frac{\ell^2}{4(4\pi^2)^3\epsilon^2}\left(\frac{1}{\delta}+\frac{6\ln\frac{\delta}{2}-1}{2}\right)\frac{1}{\abs{\un{x}-\un{y}}^4}\left(1+\ln\left(\frac{4\epsilon^2}{\abs{\un{x}-\un{y}}^2}\right)\right)
\end{align}
We see that all the quantum corrections to the two point function can be absorbed into a massshift counterterm. In quantum field theory in flat space time the common renormalization scheme requires then, that the on shell mass is fixed to be the physical mass. In this case the relevant on-shell quantity is the conformal dimension of the dual operator which is fixed to be $\Delta=2$.

\subsubsection{Four point function}
Now we can calculate the only connected one loop contribution to the second order four point function which is the diagram:
\begin{align}
	L_1(\un{x}_1,\un{x}_2,\un{x}_3,\un{x}_4)=&\begin{tikzpicture}[baseline=(base)]
		\begin{feynman}[inline=(base)]
			\vertex (a);
			\vertex [right=1.9cm of a] (b);
			\tikzfeynmanset{every vertex=dot}
			\vertex [right=0.1cm of a, label=90:$x_1$] (x1);
			\vertex [right=0.4cm of x1, label=90:$x_2$] (x2);
			\vertex [right=0.8cm of x2, label=90:$x_3$] (x3);
			\vertex [right=0.4cm of x3, label=90:$x_4$] (x4);
			\vertex [below right=0.4cm and 0.5cm of x1] (x);
			\vertex [below left=0.4cm and 0.5cm of x4] (z);
			\tikzfeynmanset{every vertex={empty dot,minimum size=0mm}}
			\vertex [below=0.2cm of x1] (base);
			\diagram* {
				(a)--[double](x1)--[double](x2)--[double](x3)--[double](x4)--[double](b),
				(x1)--[out=270, in=170, min distance=0.1cm](x)--[out=120, in=270, min distance=0.1cm](x2),
				(x3)--[out=270, in=60, min distance=0.1cm](z)--[out=10, in=270, min distance=0.1cm](x4),
				(x)--[out=80, in=100, min distance=0.3cm](z)--[out=260, in=280, min distance=0.3cm](x),
			};
		\end{feynman}
	\end{tikzpicture}=\int\dd^4x\dd^4y\sqrt{g(x)g(y)}K(\un{x}_1,x)K(\un{x}_3,x)i^2G^2(x,y)K(y,\un{x}_2)K(y,\un{x}_4)\\
	=&\int\dd^3\un{x}\dd^3\un{y}\int\limits_{-\infty}^{0}\dd\eta_x\dd\eta_y\frac{\ell^8}{(\eta_x\eta_y)^4}
	K(\un{x}_1,x)K(\un{x}_3,x)i^2G^2(x,y)K(y,\un{x}_2)K(y,\un{x}_4)
\end{align}
We have done the explicit computation in de Sitter space for all the previous diagrams and we figured out that they give the same results as the ones obtained for Euclidean Anti-de-Sitter with an additional prefactor of $-i$ for each vertex. This makes sense from the point of view that de Sitter in Poincar\'e coordinates is just the double analytic continuation $z\rightarrow-i\eta$ and $\ell_{AdS}\rightarrow -i\ell$. Therefore we can simply use the results from EAdS \cite{Bertan:2018khc} and use this continuation to get the result:
\begin{align}
	L_1(\un{x}_1,\un{x}_2,\un{x}_3,\un{x}_4)=&-\frac{3N_{\phi}^2}{16\pi^4}\left[\frac{1}{4}\psi(\delta)L_0+\frac{1}{8}\tilde{L}_1-\frac12 L_0\right]\\
	\tilde{L}_1=&\int_0^{\infty}\dd s\int_0^1\dd r\frac{sr(1-r)\log(1+s)}{(1+s)[sr(1-r)x+ry+(1-r)z]^2}\\
	\psi(\delta)=&-\frac{5}{3}-\log\frac{\delta}{8}
\end{align}
where the variables $x,y,z$ depend on the conformal invariants $u$ and $v$ depending on the scattering channel. The total four point function to second order in $\lambda$ is therefore:
\begin{align}
	\expect{\Op(\un{x}_1)\Op(\un{x}_2)\Op(\un{x}_3)\Op(\un{x}_4)}=&\frac{N_{\phi}^2}{(x_{12}x_{34})^4}\left[1+u^2+\frac{u^2}{v^2}-\frac{3\lambda}{4\pi^2}\left(\frac{u}{v}\right)^2L_0+\frac{9\lambda^2}{2\cdot 64\pi^4}\psi(\delta)\left(\frac{u}{v}\right)^2L_0\right.\\
	&\left.+\frac{3\lambda^2}{256\pi^4}\left(\frac{u}{v}\right)^2\left(\sum_{s,t,u}\tilde{L}_1-12 L_0\right)\right]+\Op(\lambda^3)
\end{align}
The UV divergence as $\delta\rightarrow0$ can be absorbed into the coupling constant in the usual way:
\begin{align}
	\lambda=\lambda_R-\frac{3\lambda_R^2}{32\pi^2}\psi(\delta)+\Op(\lambda_R)\qquad
	\Rightarrow \beta=-\fpartial{\lambda}{\ln\sqrt{\delta}}=\frac{3\lambda_R}{16\pi^2}+\Op(\lambda_R^3)
\end{align}
Now we can write the four point function in terms of the renormalized coupling constant:
\begin{align}
	\expect{\Op(\un{x}_1)\Op(\un{x}_2)\Op(\un{x}_3)\Op(\un{x}_4)}=&\frac{N_{\phi}^2}{(x_{12}x_{34})^4}\left[1+u^2+\left(\frac{u}{v}\right)^2-\frac{3\lambda_R}{4\pi^2}\left(\frac{u}{v}\right)^2L_0\right.\\
	&+\left.\frac{3\lambda_R^2}{4^4\pi^4}\left(\frac{u}{v}\right)^2\left(\sum_{s,t,u}\tilde{L}_1-12 L_0\right)\right]+\Op(\lambda_R^3)
	\label{eq:4point_full}
\end{align}

\section{Reconstruction of the dual conformal field theory}
\subsection{Conformal block expansion}
We can now look at the form of the two-point and four-point function to reconstruct the data of the dual conformal field theory. The two point function is given by:
\begin{align}
	\expect{\Op(\un{x_1})\Op(\un{x_2})}=&\frac{\delta^2\Psi[\phi_0]}{\delta\phi_0(\un{x}_1)\delta\phi_0(\un{x}_2)}
	=-\frac{N_{\phi}}{\abs{\un{x}_1-\un{x}_2}^4}
\end{align}
As $N_{\phi}>0$ the two point function is negative and therefore the conformal field theory is non-unitary.

The result for the free four point function has the form of a generalized free field. This means it is free in the sense that it is the sum over all permutations of two point functions of a free field. However there is no classical action and therefore no equation of motion which would give these two point functions, as can be checked by dimensional analysis. In theories with an internal symmetry like an $O(N)$ vector model or a matrix model, these generalized free field theories emerge when considering the effective field theory of single trace operators in the large $N$ limit.

Generalized free field theories have a very characteristic operator product expansion, which can be seen by analysing the disconnected four point function:
\begin{align}
	\expect{\Op(\un{x_1})\Op(\un{x_2})\Op(\un{x_3})\Op(\un{x_4})}=&
	\frac{N_{\phi}^2}{(x_{12}x_{34})^4}\left(1+u^2+\left(\frac{u}{v}\right)^2\right)
	=\frac{N_{\phi}^2}{(x_{12}x_{34})^4}\left(1+\left(\frac{u}{v}\right)^2\left(1+v^2\right)\right)\label{eq:4pointexp}
\end{align}
To extract the operator spectrum of the CFT from this equation, we need to expand this equation. As we will see in the next section the OPE corresponds to an expansion around $\un{x}_1\rightarrow \un{x}_2$ and $\un{x}_3\rightarrow \un{x}_4$ which in the conformal invariants is given by $u\rightarrow0$ and $v\rightarrow 1$. Expanding \eqref{eq:4pointexp} in this limit we get:
\begin{align}
	\expect{\Op(\un{x_1})\Op(\un{x_2})\Op(\un{x_3})\Op(\un{x_4})}=&
	\frac{N_{\phi}^2}{(x_{12}x_{34})^4}\left(1+\left(\frac{u}{v}\right)^2\left(2+\sum\limits_{n=1}^{\infty}(n+1)\left(1-\frac{1}{v}\right)^n\right)\right)
\end{align}
Now looking at this from the CFT perspective we can take any general four point function $\expect{\Op(\un{x_1})\Op(\un{x_2})\Op(\un{x_3})\Op(\un{x_4})}$ of operators of the same scaling dimension $\Delta_{\Op}$. Doing a double OPE between the operators $\Op(x_1)\Op(x_2)$ and $\Op(x_3)\Op(x_4)$ we get \cite{Rychkov:2016iqz}:
\begin{align}
	\expect{\Op(\un{x_1})\Op(\un{x_2})\Op(\un{x_3})\Op(\un{x_4})}=&\sum\limits_{\tilde{\Op}\tilde{\tilde{\Op}}}\lambda_{\tilde{\Op}}\lambda_{\tilde{\tilde{\Op}}}A_{\tilde{\Op}}(x_{12},\partial_2)A_{\tilde{\tilde{\Op}}}(x_{34},\partial_4)\expect{\tilde{\Op}(x_2)\tilde{\tilde{\Op}}(x_4)}\\
	&=\sum\limits_{\tilde{\Op}}\lambda^2_{\tilde{\Op}}A_{\tilde{\Op}}(x_{12},\partial_2)A_{\tilde{\Op}}(x_{34},\partial_4)\frac{1}{x_{24}^{2\Delta}}\\
	&=\frac{1}{(x_{12}x_{34})^{2\Delta_{\Op}}}\sum\limits_{\tilde{\Op}}\lambda_{\tilde{\Op}}^2G_{\tilde{\Op}}(u,v)
\end{align}
where $G_{\tilde{\Op}}(u,v)$ are the conformal blocks with conformal dimension $\tilde{\Delta}$ and spin $l$, $\lambda_{\tilde{\Op}}$ are the OPE coefficients und $u$ and $v$ are the conformal invariants.

To see which operators show up in the OPE of \eqref{eq:4pointexp} we look at certain class of operators which can be constructed by combining two single trace primary operators into a new double trace primary operator $:\Op^2:_{n,l}=:\Op\square^n\partial^l\Op:$ with spin $l$ and scaling dimension $\Delta(n,l)=2\Delta_{\Op}+2n+l$. Using a conglomeration mechanism we can calculate the OPE coefficients of these conformal blocks, which was done in \cite{Fitzpatrick:2011dm} and get the result for $\Delta_{\Op}=2$:
\begin{align}
	\lambda_{:\Op:_{n,l}}^2=C_{\Delta(n,l),l}&=\frac{2^{-l-4n}\Gamma\left(l+\frac32\right)\Gamma\left(n+\frac32\right)\Gamma(n+2)\Gamma(l+n+2)\Gamma\left(l+n+\frac52\right)\Gamma(l+2n+3)}
	{\Gamma(l+1)\Gamma(n+1)^2\Gamma\left(l+n+\frac32\right)^2\Gamma\left(l+2n+\frac52\right)}
\end{align}
To find the form of the actual conformal blocks we use the fact that the conformal blocks correspond to eigenstates $\ket{\Delta,l}$ of the quadratic casimir $C$ of the conformal group:
\begin{align}
	C&=-\frac12J^{ab}J_{ab}=D(D-d)-\frac12M^{\mu\nu}M_{\mu\nu}\\
	\Rightarrow C\ket{\Delta,l}&=\left[\Delta(\Delta-d)+l(l+d-2)\right]\ket{\Delta,l}
\end{align}
The differential operator $\mathcal{D}$ representing the action of the quadratic casimir on the conformal block can be written in terms of $u$ and $v$ and so the conformal blocks have to fulfill the differential equation:
\begin{align}
	\left[-\left((1-v)^2-u(1+v)\right)\fpartial{}{v}v\fpartial{}{v}-(1-u+v)u\fpartial{}{u}u\fpartial{}{u}\right.&\left.+2(1+u-v)uv\frac{\partial^2}{\partial u\partial v}\right]
	G_{\Delta,l}(u,v)=\\
	&=-\left[\Delta(\Delta-d)+l(l+d-2)\right]G_{\Delta,l}(u,v)
\end{align}
The solution to this equation for $l=0$ is given by:
\begin{align}
	&\Rightarrow G_{\Delta,0}(u,v)=\left(\frac{u}{v}\right)^{\frac{\Delta}{2}}\sum\limits_{m,n=0}^{\infty}
	\frac{\left(\frac{\Delta}{2}\right)_m^2\left(\frac{\Delta}{2}\right)_{m+n}^2}{m!n!(\Delta+1-d/2)_m(\Delta)_{2m+n}}\left(\frac{u}{v}\right)^m\left(1-\frac{1}{v}\right)^n\\
	&\text{with the Pochhammer symbol defined as: }(x)_n=\frac{\Gamma(x+n)}{\Gamma(x)}\label{eq:Pochhammer}
\end{align}
The conformal blocks with $l>0$ can be obtained by a complicated iteration relation. This relation together with the details of the above calculation can be found in \cite{Dolan:2003hv,Dolan:2000ut}.

We can now compare the conformal blocks together with the OPE coefficients to see which double trace operators show up in the expansion of our four point function \eqref{eq:4pointexp}:
\begin{align}
	\expect{\Op(\un{x_1})\Op(\un{x_2})\Op(\un{x_3})\Op(\un{x_4})}=&
	\frac{1}{(x_{12}x_{34})^{2\Delta_{\Op}}}\left(1+\sum\limits_{n,l}C_{\Delta(n,l),l}G_{\Delta(n,l),l}(u,v)\right)\label{eq:4point_ope}
\end{align}
By comparing \eqref{eq:4point_ope} to \eqref{eq:4pointexp} we find that the spectrum of the double trace operators contains all $n\geq 0$ but only even spin contributions.

\subsection{Anomalous dimensions from interactions}
We now have the conformal block expansion of the free scalar field theory at $\eta\rightarrow0$. It consists of an infinite sum of double trace operators $:\Op\square^n\partial^l\Op:$ with all integer $n\geq0$ and all even $l\geq0$.

Next we want to incorporate the interaction terms we calculated in the previous chapter into the conformal block expansion. This happens in a similar way as in AdS, which is described for example in \cite{Fitzpatrick:2010zm}, where the interaction generates a shift in the scaling dimension of the double trace operators. To calculate these anomalous dimensions we look at a general conformal block expansion of a four point function in conformal blocks of dimension $\Delta$ and spin $l$:
\begin{align}
	\expect{\Op_{\Delta}(x_1)\Op_{\Delta}(x_2)\Op_{\Delta}(x_3)\Op_{\Delta}(x_4)}&=\frac{1}{(x_{12}x_{34})^{2\Delta}}\left(1+\sum\limits_{\Delta,l}\mathfrak{C}_{\Delta,l}\mathfrak{G}_{\Delta,l}\right)
\end{align}
where $\mathfrak{C}_{\Delta,l}$ and $\mathfrak{G}_{\Delta,l}$ are the generalized OPE coefficients and conformal blocks for general scaling dimensions.

By introducing interactions in the bulk theory we expect the operators to collect anomalous dimensions like:
\begin{align}
	\tilde{\Delta}_{n,l}&=\Delta_{n,l}+\gamma^{(1)}_{n,l}+\gamma^{(2)}_{n,l}+...
\end{align}
where it is assumed that $\gamma^{(i)}_{n,l}\propto\lambda_R^{i}$, where $\lambda_R$ is the renormalized coupling constant in the bulk theory. Expanding the OPE coefficients and the conformal blocks to second order in the coupling constant gives us:
\begin{align}
	\mathfrak{C}_{\Delta(n,l),l}=&C_{\Delta(n,l),l}+(\gamma^{(1)}_{n,l}+\gamma^{(2)}_{n,l})C^{(1)}_{n,l}+\frac12(\gamma^{(1)}_{n,l})^2C^{(2)}_{n,l}+...\\
	\mathfrak{G}_{\Delta(n,l),l}=&G_{\Delta(n,l),l}+(\gamma^{(1)}_{n,l}+\gamma^{(2)}_{n,l})\underbrace{\left.\fpartial{G_{\Delta,l}}{\Delta}\right\vert_{\Delta(n,l)}}_{G'_{\Delta(n,l),l}}
	+\frac12(\gamma^{(1)}_{n,l})^2\underbrace{\left.\fnpartial{2}{G_{\Delta,l}}{\Delta}\right\vert_{\Delta(n,l)}}_{G''_{\Delta(n,l),l}}+...\\
	\Rightarrow \mathfrak{C}_{\Delta(n,l),l}\mathfrak{G}_{\Delta(n,l),l}=&C_{\Delta(n,l),l}G_{\Delta(n,l),l}+\gamma^{(1)}_{n,l}\left(C_{\Delta(n,l),l}G'_{\Delta(n,l),l}+C^{(1)}_{n,l}G_{\Delta(n,l),l}\right)\\
	&+\frac{1}{2}(\gamma^{(1)}_{n,l})^2\left(C_{\Delta(n,l),l}G''_{\Delta(n,l),l}+C^{(2)}_{n,l}G_{\Delta(n,l),l}+2C^{(1)}_{\Delta(n,l),l}G'_{\Delta(n,l),l}\right)+...
\end{align}
And therefore the expansion of the four point function to second order in the bulk coupling constant is given by:
\begin{align}
	&\expect{\Op_{\Delta}(x_1)\Op_{\Delta}(x_2)\Op_{\Delta}(x_3)\Op_{\Delta}(x_4)}=\frac{1}{(x_{12}x_{34})^{2\Delta}}\left(1+\sum\limits_{n,l}C_{\Delta(n,l),l}G_{\Delta(n,l),l}\right.\\
	&+\sum\limits_{n,l}\gamma^{(1)}_{n,l}\left(C_{\Delta(n,l),l}G'_{\Delta(n,l),l}+C^{(1)}_{n,l}G_{\Delta(n,l),l}\right)\\
	&+\sum\limits_{n,l}\gamma^{(2)}_{n,l}\left(C_{\Delta(n,l),l}G'_{\Delta(n,l),l}+C^{(1)}_{n,l}G_{\Delta(n,l),l}\right)\\
	&\left.+\sum\limits_{n,l}\frac{1}{2}(\gamma^{(1)}_{n,l})^2\left(C_{\Delta(n,l),l}G''_{\Delta(n,l),l}+C^{(2)}_{n,l}G_{\Delta(n,l),l}+2C^{(1)}_{\Delta(n,l),l}G'_{\Delta(n,l),l}\right) \right)\label{eq:4point_anom}
\end{align}
Now we can calculate for each order in $\lambda_R$ the expansion in $u/v$ and $1-v^{-1}$ and compare it to the bulk calculation to get the anomalous dimensions and OPE coefficients.

The boundary to boundary four point contribution to the wave function that we calculated in the previous chapter can be written in the form:
\begin{align}
	\expect{\Op(x_1)\Op(x_2)\Op(x_3)\Op(x_4)}=&\frac{\Nphi^2}{(x_{12}x_{34})}\left(1+\left(\frac{u}{v}\right)^2\sum_{m,n=0}^{\infty}\mathcal{F}_{mn}(\lambda_R,\log(u/v))(1-v^{-1})^n\left(\frac{u}{v}\right)^m\right)\label{eq:4point_Fmn}
\end{align}
where the coefficients of the expansion $\mathcal{F}_{mn}(\lambda_R,\log(u/v))$ can be read off from \eqref{eq:4point_full}.

The first five coefficients are given by:
\begin{align}
	\mathcal{F}_{00}&=\Nphi^2\left(2+\frac{\lambda_R}{48\pi^2}\left(1+3\log\frac{u}{v}\right)+\frac{\lambda_R^2}{768\pi^4}\left(5+\frac{11}{2}\log\frac{u}{v}+\frac34\left(\log\frac{u}{v}\right)^2\right)\right)\label{eq:F00}\\
	\mathcal{F}_{01}&=\Nphi^2\left(2+\frac{\lambda_R}{96\pi^2}\left(5+6\log\frac{u}{v}\right)+\frac{\lambda_R^2}{3072\pi^4}\left(31+25\log\frac{u}{v}+3\left(\log\frac{u}{v}\right)^2\right)\right)\label{eq:F01}\\
	\mathcal{F}_{10}&=\Nphi^2\left(-\frac{\lambda_R}{120\pi^2}\left(\frac{17}{5}-6\log\frac{u}{v}\right)+\frac{\lambda_R^2}{38400\pi^4}\left(-\frac{491}{5}+292\log\frac{u}{v}+30\left(\log\frac{u}{v}\right)^2\right)\right)\label{eq:F10}\\
	\mathcal{F}_{11}&=N^2_{\phi}\left(-\frac{\lambda_R}{40\pi^2}\left(\frac{7}{10}-3\log\frac{u}{v}\right)
	+\frac{\lambda_R^2}{38400\pi^4}\left(-\frac{13}{10}+468\log\frac{u}{v}+45\left(\log\frac{u}{v}\right)^2\right)\right)\label{eq:F11}\\
	\mathcal{F}_{02}&=\Nphi^2\left(3+\frac{9\lambda_R}{160\pi^2}\left(\frac{11}{10}+\log\frac{u}{v}\right)+\frac{\lambda_R^2}{256000\pi^4}\left(2816+1965\log\frac{u}{v}+225\left(\log\frac{u}{v}\right)^2\right)\right)\label{eq:F02}
\end{align}
By comparing this to the conformal block expansion in \eqref{eq:4point_anom}, we can solve it order by order for the anomalous dimensions and conformal blocks.

The only contribution to the $00$ component comes from the $n=0,l=0$ part of the conformal block expansion:
\begin{align}
	\mathfrak{C}_{\Delta(0,0),0}\mathfrak{G}_{\Delta(0,0),0}=&
	\left(\frac{u}{v}\right)^2\left[2+\gamma^{(1)}_{0,0}\left(\log\frac{u}{v}+C^{(1)}_{0,0}\right)
	\gamma^{(2)}_{0,0}C^{(1)}_{0,0}+\frac12(\gamma^{(1)}_{0,0})^2C^{(2)}_{0,0}\right.\\
	&+\frac12\log\frac{u}{v}\left(2\gamma^{(2)}_{0,0}+(\gamma^{(1)}_{0,0})^2C^{(1)}_{0,0}\right)
	\left.+\frac14\left(\log\frac{u}{v}\right)^2(\gamma^{(1)}_{0,0})^2\right]+\Op(1-v^{-1},u/v)
\end{align}
So by looking at the part which is first order in $\lambda_R$ in equation \eqref{eq:F00} we can solve it to get the first order contributions:
\begin{align}
	\gamma^{(1)}_{0,0}=\frac{\lambda_R}{16\pi^2};\ \ \ C^{(1)}_{0,0}=\frac13
\end{align}
Going to the next order in $\lambda_R$ we can repeat this procedure to get the second order contributions to the anomalous dimensions and the OPE coefficients:
\begin{align}
	\gamma^{(1)}_{0,0}=\pm\frac{\lambda_R}{16\pi^2};\ \ \ \gamma^{(2)}_{0,0}=\frac{5\lambda_R^2}{768\pi^4}=\frac53\gamma^{(1)}_{0,0};\ \ \ C^{(2)}_{0,0}=\frac{20}{9}		
\end{align}
Note that in the same way as in \cite{Bertan:2018afl} the first order anomalous dimension coincides at different order in $\lambda_R$ contributions to the conformal block expansion.

Now we plug these results back into the conformal block expansion and then solve for the next contribution in the $(1-v^{-1}),\frac{u}{v}$ expansion.

We then recognize that the $01$ contribution of the $n=0,l=0$ conformal block with the above results plugged in is equal to \eqref{eq:F01}. Therefore any other conformal block contribution with a term of order $01$ must be zero and so the anomalous dimension and OPE coefficient from that block must be zero. This is the case for $n=0,l=1$ and so the spin 1 contributions vanishes. We will find that this happens for all odd spin contributinos. So we see that the interaction just shifts the scaling dimension of the conformal blocks compared to the generalized free theory and does not generate new contributions in the conformal block expansion.

Since the spin 1 contribution is zero and the $00$ contribution is completely determined by $n=0,l=0$ conformal block, the next order in the expansion with only one more contribution is the $02$ term to which the $n=0,l=0$ and the $n=0,l=2$ block contribute. After plugging in the results we already obtained for $n=0,l=0$ the sum of those two terms is given by:
\begin{align}
	&\frac{9\lambda_R}{160\pi^2}\left(\frac{11}{10}+\log\frac{u}{v}\right)
	+\frac{1}{20}\gamma^{(1)}_{0,2}\left(12\log\frac{u}{v}+5C^{(1)}_{0,2}\right)
	+\frac{\lambda_R^2}{10240\pi^4}\left(\frac{2827}{25}+\frac{399}{5}\log\frac{u}{v}+9\left(\log\frac{u}{v}\right)^2\right)\\
	&+\frac{1}{40}\left(10\gamma^{(2)}_{0,2}C^{(1)}_{0,2}+5(\gamma^{(1)}_{0,2})^2C^{(2)}_{0,2}+
	\log\frac{u}{v}\left(24\gamma^{(2)}_{0,2}+5(\gamma^{(1)}_{0,2})^2C^{(2)}_{0,2}\right)
	+6\left(\log\frac{u}{v}\right)^2(\gamma^{(1)}_{0,2})^2\right)
\end{align}
Comparing this to \eqref{eq:F02} we can solve it to get:
\begin{align}
	\gamma^{(1)}_{0,2}=0;\ \ \ \gamma^{(2)}_{0,2}=-\frac{\lambda_R^2}{5120\pi^4}=-\frac{1}{20}\gamma^{(1)}_{0,0};\ \ \ 
	C^{(1)}_{0,2}=\frac{22}{25}
\end{align}
Again the values for $\gamma^{(1)}_{0,2}$ coincide at different orders in $\lambda_R$.

By repeating this procedure for more orders we find the following general procedure to calculate the OPE coefficients and anomalous dimensions:
\begin{enumerate}[$\bullet$]
	\item No new conformal blocks that did not appear in the expansion of the generalized free theory are generated by the interaction, i.e. only blocks with even $l$ contribute.
	\item If we denote the expansion coefficient in the $(u/v)^m,(1-v^{-1})^k$ expansion of the conformal block $\mathfrak{C}_{\Delta(n,l),l}\mathfrak{G}_{\Delta(n,l),l}$ by $\left[\mathfrak{C}_{\Delta(n,l),l}\mathfrak{G}_{\Delta(n,l),l}\right]_{mk}$, then the coefficients $\mathcal{F}_{mk}$ in \eqref{eq:4point_Fmn} are given by:
	\begin{align}
		\mathcal{F}_{mk}=\sum\limits_{i=0}^m\sum\limits_{j=0(\text{even})}^{2n+k}
		\left[\mathfrak{C}_{\Delta(i,j),j}\mathfrak{G}_{\Delta(i,j),j}\right]_{mk}\label{eq:algorithm}
	\end{align}
	\item Equation \eqref{eq:algorithm} can now be solved iteratively. We start with the lowest order and solve it for $\gamma^{(1)}_{n,l}, \gamma^{(2)}_{n,l}$ and $C^{(1)}_{n,l}$, plug the results back into $\mathfrak{C}_{\Delta(n,l),l}\mathfrak{G}_{\Delta(n,l),l}$ and then go to the next order.
\end{enumerate}
Going through this algorithm we obtain the following results.

From the first order in $\lambda_R$ we get:
\begin{align}
	\gamma^{(1)}_{n,0}&=\frac{\lambda_R}{16\pi^2}\qquad \gamma^{(1)}_{n,l\neq 0}=0\\
	C^{(1)}_{0,0}&=\frac13\qquad C^{(1)}_{1,0}=-\frac{478}{735}\qquad 
	C^{(1)}_{2,0}=-\frac{11507}{50820}\qquad C^{(1)}_{3,0}=-\frac{278092}{6441435}\qquad C^{(1)}_{4,0}=-\frac{609121355}{97749595944}\\
	\Rightarrow& C^{(1)}_{n,0}=\frac12\fpartial{C_{\Delta(n,0),0}}{n} 
\end{align}
For later convenience we define:
\begin{align}
		\tilde{\gamma}^{(2)}_{n,l}\equiv&\gamma_{n,l}^{(2)}\times\frac{256\pi^4}{\lambda_R^2}
\end{align}
Repeating the above procedure for higher spin up to $l=18$ and $n=9$ we find the following data from the second order in $\lambda_R$ contributions (a complete list of these anomalous dimensions and first order OPE coefficients can be found in appendix B in table \ref{tab:OPE1} and \ref{tab:gammanl} and the second order OPE coefficients are given in table \ref{tab:OPE2}):
\begin{tiny}
\begin{align}
	\tilde{\gamma}^{(2)}_{0,0}=&\frac{5}{3}; && \tilde{\gamma}^{(2)}_{0,2}=-\frac{1}{20}; && \tilde{\gamma}^{(2)}_{0,4}=-\frac{1}{140}; && \tilde{\gamma}^{(2)}_{0,6}=-\frac{1}{504}; && \tilde{\gamma}^{(2)}_{0,8}=-\frac{1}{1320}; && \tilde{\gamma}^{(2)}_{0,10}=-\frac{1}{2860};\\
	\tilde{\gamma}^{(2)}_{1,0}=&\frac{46}{15}; && \tilde{\gamma}^{(2)}_{1,2}=-\frac{107}{1260}; && \tilde{\gamma}^{(2)}_{1,4}=-\frac{19}{1260}; && \tilde{\gamma}^{(2)}_{1,6}=-\frac{131}{27720}; && \tilde{\gamma}^{(2)}_{1,8}=-\frac{301}{154440}; && \tilde{\gamma}^{(2)}_{1,10}=-\frac{19}{20020};\\
	\tilde{\gamma}^{(2)}_{2,0}=&\frac{113}{28}; && \tilde{\gamma}^{(2)}_{2,2}=-\frac{269}{2520}; && \tilde{\gamma}^{(2)}_{2,4}=-\frac{6707}{311850}; && \tilde{\gamma}^{(2)}_{2,6}=-\frac{1973}{270270} && \tilde{\gamma}^{(2)}_{2,8}=-\frac{3439}{1081080}; && \tilde{\gamma}^{(2)}_{2,10}=-\frac{59281}{36756720};\\
	\tilde{\gamma}^{(2)}_{3,0}=&\frac{535}{112}; && \tilde{\gamma}^{(2)}_{3,2}=-\frac{6697}{55440}; && \tilde{\gamma}^{(2)}_{3,4}=-\frac{19037}{720720}; && \tilde{\gamma}^{(2)}_{3,6}=-\frac{143581}{15135120}; && \tilde{\gamma}^{(2)}_{3,8}=-\frac{175939}{40840800}; && \tilde{\gamma}^{(2)}_{3,10}=-\frac{349787}{155195040};\\
	\tilde{\gamma}^{(2)}_{4,0}=&\frac{745007}{138600}; && \tilde{\gamma}^{(2)}_{4,2}=-\frac{880571}{6756750} && \tilde{\gamma}^{(2)}_{4,4}=-\frac{13589}{450450}; && \tilde{\gamma}^{(2)}_{4,6}=-\frac{138443}{12252240}; && \tilde{\gamma}^{(2)}_{4,8}=-\frac{2775023}{523783260}; && \tilde{\gamma}^{(2)}_{4,10}=-\frac{1654313}{581981400};\\
\end{align} 
\end{tiny}
\begin{table}
\begin{center}
\begin{tabular}{|c"c|}
	\hline 
	$n \setminus l$ & $0$ \\
	\thickhline
	$0$ & $\frac{20}{9}$\\
	\hline
	$1$ & $\frac{111392}{77175}$\\
	\hline
	$2$ & $\frac{27588119}{70436520}$\\
	\hline
	$3$ & $\frac{6664117739}{96718146525}$\\
	\hline
	$4$ & $\frac{54416659121622349}{5688844669692344160}$\\
	\hline
	$5$ & $\frac{8887970348158281431}{7764634802452359932100}$\\
	\hline
	$6$ & $\frac{9527783274829256957833}{77012500285547896469400000}$\\
	\hline
	$7$ & $\frac{8939233385397483916814706947}{720011509787613310332374452893750}$\\
	\hline
	$8$ & $\frac{502743179123371126837198189}{426950728014025268129610534000000}$\\
	\hline
	$9$ & $\frac{497605048296141707730882676711}{4658017703129473554739907996885428500}$\\
	\hline
\end{tabular}
\caption{OPE coefficients $C_{n,0}^{(2)}$}
\label{tab:OPE2}
\end{center}
\end{table}
For the second order anomalous dimensions we can guess the following general structure for $l>0$:
\begin{align}
	\tilde{\gamma}^{(2)}_{n,l}=\frac{1}{(l)_{4+2n}}\sum\limits_{m=0}^n\afrk{n}{n-m}(l+2+m)_{2(n-m)}\label{eq:gammanl}
\end{align}
where $\mathfrak{a}^{(n)}_{n-m}$ are free parameters and $(l+2+m)_{2(n-m)}$ are the Pochhammer symbols given in \eqref{eq:Pochhammer}. The $\mathfrak{a}^{(n)}_{n-m}$s can be fixed by solving the system of linear equations with the anomalous dimensions we calculated above. For each $n$ we need $n+1$ known results to solve the equations. With the data we have, we can determine the $\mathfrak{a}^{(n)}_{n-m}$s up to $n=4$:
\begin{align}
	\afrk{0}{0}=&-6;\\
	\afrk{1}{0}=&12;&& \afrk{1}{1}=-22;\\
	\afrk{2}{0}=&-96;&& \afrk{2}{1}=\frac{224}{3};&& \afrk{2}{2}=-\frac{146}{3};\\
	\afrk{3}{0}=&1440;&& \afrk{3}{1}=-1080;&& \afrk{3}{2}=252;&& \afrk{3}{3}=-86;\\
	\afrk{4}{0}=&-34560;&& \afrk{4}{1}=25344;&& \afrk{4}{2}=-\frac{28512}{5};&& \afrk{4}{3}=\frac{3168}{5};&& \afrk{4}{4}=-134;
\end{align}
So the solutions for $\gamma^{(2)}_{n,l}$ up to $n=4$ for $l>0$ are given by:
\begin{align}
	\tilde{\gamma}^{(2)}_{0,l}=&-\frac{6}{(l)_4}\label{eq:nzero}\\
	\tilde{\gamma}^{(2)}_{1,l}=&-\frac{2}{(l)_6}(11(l+2)_2-6)\label{eq:none}\\
	\tilde{\gamma}^{(2)}_{2,l}=&-\frac{2}{3(l)_8}(73(l+2)_4-112(l+3)_2+144)\label{eq:ntwo}\\
	\tilde{\gamma}^{(2)}_{3,l}=&-\frac{2}{(l)_{10}}(43(l+2)_6-126(l+3)_4+540(l+4)_2-720)\label{eq:nthree}\\
	\tilde{\gamma}^{(2)}_{4,l}=&-\frac{2}{5(l)_{12}}(335(l+2)_8-1584(l+3)_6+14256(l+4)_4-63360(l+5)_2+86400)\label{eq:nfour}
\end{align}
The equation for $n=0$ given in \eqref{eq:nzero} was already discovered in \cite{Bertan:2018afl,Bertan:2018khc}. We now have reduced the problem of finding the anomalous dimensions of arbitrary spin for each $n$ to solving a system of $n+1$ linear equations.

Note, however, that while equations \eqref{eq:nzero}, \eqref{eq:none}, \eqref{eq:ntwo}, \eqref{eq:nthree} and \eqref{eq:nfour} seem to be exact and correctly predict higher $l$ contributions which were not used to find these equations, the general form of \eqref{eq:gammanl} should only be treated as an educated guess by now and is in no way mathematically proven for general $n$. To confirm the form of \eqref{eq:gammanl} for higher $n$ we would need to compute more anomalous dimensions using the algorithm described above. For the moment at least we unfortunately are restricted from proceeding by limitation of computational resources.

We have not been able to find similar relations for the OPE coefficients. The OPE coefficients at first order in $\lambda$ can be found in the Appendix B. 

\section{Conclusions}
Let us compare the results we got to the ones obtained from calculations in EAdS in \cite{Bertan:2018khc,Bertan:2018afl} and to similar calculations done in dS (see e.g. \cite{Anninos:2014lwa,Maldacena:2002vr}).

We see that the action of the scalar field in de Sitter given in \eqref{eq:action} transforms in the following way under the continuation from dS to EAdS ($\eta=iz$, $\ell=i\ell_{AdS}$):
\begin{align}
	S_{dS}&\longrightarrow \frac{i\ell^2_{AdS}}{2}\int\limits_0^{\infty}\frac{\dd z\dd^3x}{z^4}\left\{z^2(\partial_z\phi)^2+z^2(\nabla\phi)^2+m^2\ell_{AdS}^2\phi^2+\frac{\lambda\ell_{AdS}^2}{12}\phi^4\right\}=iS_{EAdS}\label{eq:actionEAdS}
\end{align}
Therefore we have $iS_{dS}\rightarrow-S_{EAdS}$ which is the original motivation for \eqref{eq:partitionfunction} which relates the partition function in EAdS to the Bunch-Davies wave function in dS. We found that this relation holds not only at the tree-level but also to second order in quantum corrections, which was explicitly checked for the two point function. 

Since $\ell\rightarrow i\ell$ we can see from \eqref{eq:on_shell_2pnt} that the two point function just switches the sign. However this new sign can be absorbed into the normalization constant. After this normalization is fixed, the sign of all the other higher order functions is fixed as well and we see that the signs match exactly with the EAdS calculation and we come to the conclusion that all the loop corrections of the two point function are proportional to the mass shift and therefore can be absorbed into the scaling dimension on the boundary.

In case of the four point function we find that the results again coincide exactly with the calculations in EAdS. The interactions in the bulk generate anomalous dimensions of the double trace operators of the dual CFT. We calculated those by comparing the conformal block expansion of a deformed generalized free field with the $\eta\rightarrow0$ limit of the four point function as it was done in \cite{Bertan:2018afl}. Unsurprisingly we therefore find that the anomalous dimensions and OPE coefficients coincide up to a sign in the first order which is due to a different sign of the coupling constant in \cite{Bertan:2018afl}.

We found a general structure in the anomalous dimensions in \eqref{eq:gammanl} which generalizes the result from \cite{Bertan:2018afl,Bertan:2018khc}. The result is in agreement with general results in \cite{Basso:2006nk,Fitzpatrick:2012yx,Alday:2015eya}.

Next we would like to comment on the potential impact of these results for the conjectured higher spin/CFT duality. In \cite{Anninos:2011ui} a duality between an interacting $Sp(N)$ model and Vasiliev higher spin theory in dS was conjectured in analogy to the duality between an interacting $O(N)$ model and higher spin theory in AdS. Going from $O(N)$ to $Sp(N)$ corresponds to continuing $N\rightarrow-N$. As $N$ is the parameter controlling the quantum fluctuations in the $1/N$ expansion it is the prefactor of the action in the partition function. This would mean that the action should switch the sign and it was argued in \cite{Anninos:2011ui} that as a consequence every connected tree level $n$ point function would acquire an extra minus sign when going from EAdS to dS. By looking at the analysis performed in section 3 we see that the two point function does in fact change sign while the four point function does not. For the two point function this is just a normalization issue as mentioned above. However after this normalization is fixed the relative sign of all the higher order contributions to the $n$ point functions are fixed and we get the same result as for EAdS, i.e. no change of sign in the four point function. In order to have a sign change one has to additionally change the sign of the $\phi^4$ coupling.


Another way to see this is by looking at the Euclidean action in AdS with a positive coupling constant \eqref{eq:actionEAdS}. After a field redefinition $\tilde{\phi}=\sqrt{\lambda}\phi$ the action reads:
\begin{align}
	S_{EAdS}=&\frac{\ell^2_{AdS}}{2\lambda}\int\limits_0^{\infty}\frac{\dd z\dd^3x}{z^4}\left\{z^2(\partial_z\tilde{\phi})^2+z^2(\nabla\tilde{\phi})^2+m^2\ell_{AdS}^2\tilde{\phi}^2+\frac{\ell_{AdS}^2}{12}\tilde{\phi}^4\right\}\label{eq:actionEAdSrescaled}
\end{align}
Doing the double analytic continuation to dS by setting $\ell_{AdS}=-i\ell$ and $z=-i\eta$ we get back $-iS_{dS}$ in terms of $\tilde{\phi}$. But from equation \eqref{eq:actionEAdSrescaled} it already becomes clear that changing the sign of the action corresponds to going from $\lambda\rightarrow-\lambda$. This means that going from AdS to dS in the way described in \cite{Anninos:2011ui} would correspond to going from a globally stable potential to a globally unstable potential for a scalar field. 

However, this is no statement about the higher spin theory, since we do not know the exact form of the quartic interaction term for higher spin fields. Our analysis in this paper does not consider higher spin fields and we only consider quartic self-interactions. Thus the only implication we can make is that if such a quartic scalar interaction were present in the higher spin theory it would change sign when passing from EAdS to dS. 


\section*{Acknowledgements}
We would like to thank Igor Bertan and Evgeny Skvortsov for helpful discussions and for providing parts of the Mathematica code used in this calculation. T.H.'s work was supported by the PhD scholarship program of the Hans-B\"ockler-Stiftung. This work was supported, in parts, by the DFG Excellence cluster ORIGINS.

\newpage
\appendix
\renewcommand{\thesection}{\Alph{section}}
\numberwithin{equation}{section}
\section{Calculation of the tree-level four point function}
Here we would like to give the full calculation of the first order contribution to the four point function. We start from equation \eqref{eq:4pointSchwinger}:
\begin{align}
	M_4=&\int\limits_{\infty}^{0}\dd\eta\dd^3 x\eta^4\int\limits_0^{\infty}\prod\limits_{i=1}^{4}\dd\alpha_i\alpha_i\eul^{-i\alpha_i(\eta^2-(\un{x}_i-\un{x})^2-i\epsilon)}
\end{align}
Now the integration over $x$ and $\eta$ can be performed by rearranging the exponent to do a Gaussian integration. To make the integrals well definded we have to tilt it slightly into the complex plane so the integration boundaries are in fact $0$ and $\infty(1+i\delta)$:
\begin{align}
	M_4=&\int\limits_{-\infty}^0\dd\eta\dd^3 x\eta^4\int\limits_0^{\infty}\prod\limits_{i=1}^{4}\dd\alpha_i\alpha_i
	\exp(-i\alpha_1(\eta^2-\un{x}_1^2-i\epsilon)-i\alpha_2(\eta^2-\un{x}_2^2-i\epsilon)-i\alpha_3(\eta^2-\un{x}_3^2-i\epsilon)\\
	&-i\alpha_4(\eta^2-\un{x}_4^2-i\epsilon)+i\underbrace{(\alpha_1+\alpha_2+\alpha_3+\alpha_4)}_{:=\beta}\un{x}^2-2i\underbrace{(\alpha_1 \un{x}_1+\alpha_2 \un{x}_2+\alpha_3 \un{x}_3+\alpha_4 \un{x}_4)}_{:=\un{y}})\\
	=&\int\limits_{-\infty}^0\dd\eta\dd^3 x\eta^4\int\limits_0^{\infty}\prod\limits_{i=1}^{4}\dd\alpha_i\alpha_i
	\exp\left(-i\sum_{i=1}^{4}\alpha_i(\eta^2-\un{x}_i^2-i\epsilon)+i\beta\left(\un{x}-\frac{\un{y}}{\beta}\right)^2-i\frac{\un{y}^2}{\beta}\right)\\
	=&\left(\frac{i\pi}{\beta}\right)^{3/2}\int\limits_{-\infty}^0\dd\eta\eta^4\int\limits_0^{\infty}\prod\limits_{i=1}^{4}\dd\alpha_i\alpha_i
	\exp\left(-i\beta(\eta^2-i\epsilon)+\frac{i}{\beta}\left(\sum_{i<j=1}^{4}\alpha_i\alpha_j x_{ij}^2\right)\right)\\
	=&\frac{-3i}{8}\pi^2\int\limits_0^{\infty}\prod_{i=1}^{4}\dd\alpha_i\alpha_i\frac{1}{\beta^4}\exp\left(-\frac{1}{i\beta}\left(\sum_{i<j=1}^4\alpha_i\alpha_j x_{ij}^2\right)\right)\label{eq:feynman_par}
\end{align}
Starting form (\ref{eq:feynman_par}) we can use the invariance of this integral under $\sum_i\alpha_i\rightarrow\sum_i\xi_i\alpha_i$ if $\xi_i\geq 0$ and $\sum_i\xi_i^2\geq 0$ to write \cite{Symanzik:1972wj}:
\begin{align}
	M_4=&\frac{-3i\pi^2}{8}\int\limits_0^{\infty}\prod_{i=1}^{4}\dd\alpha_i\alpha_i\frac{1}{\left(\sum_i\xi_i\alpha_i\right)^4}
	\exp\left(-\frac{1}{i\sum_i\xi_i\alpha_i}\left(\sum_{i<j=1}^4\alpha_i\alpha_j x_{ij}^2\right)\right)
\end{align}
If we chose $\xi_i=\delta_{1i}$ then the integral simplifies to:
\begin{align}
	M_4=&\frac{-3i\pi^2}{8}\int\limits_0^{\infty}\prod_{i=1}^{4}\dd\alpha_i\alpha_i\frac{1}{\alpha_1^4}
	\exp\left(-\frac{1}{i\alpha_1}\left(\sum_{i<j=1}^4\alpha_i\alpha_j x_{ij}^2\right)\right)\\
	=&\frac{-3i\pi^2}{8(2\pi i)^2}\iint\limits_{c-i\infty}^{c+i\infty}\dd s\dd t\Gamma(s)\Gamma(t)\Gamma(2-s-t)\\
	&\times\frac{x_{23}^{2(s+t-2)}}{x_{24}^{2s}x_{34}^{2t}}
	\int\limits_0^{\infty}\prod_{i=2}^4\dd\alpha_i\alpha_2^{t-1}\alpha_3^{s-1}\alpha_4^{-s-t+1}\eul^{i\alpha_2x_{12}^2+i\alpha_3 x_{13}^2+i\alpha_4 x_{14}^2}
\end{align}
Here we used the inverse Mellin transformation of the exponential which also holds for complex arguments $x$ and is given by:
\begin{align}
	\eul^{-x}&=\frac{1}{2\pi i}\int\limits_{c-i\infty}^{c+i\infty}\Gamma(s)x^{-s}\dd s
\end{align}
where $c$ is an arbitrary positive constant. The integration over $\alpha_1$ was performed by using the definition of the gamma function. All the prefactors of $-i$ cancel out and the integrals over the remaining $\alpha_i$ can now also be rewritten as gamma functions and we arrive at
\begin{align}
	M_4=&\frac{-3i\pi^2}{8(2\pi i)^2}\frac{1}{(x_{14}x_{23})^4}\iint\limits_{c-i\infty}^{c+i\infty}\dd s\dd t\frac{(x_{14}x_{23})^{2(s+t)}}{(x_{12}x_{34})^{2t}(x_{13}x_{24})^{2s}}\Gamma(s)^2\Gamma(t)^2
	\Gamma(2-s-t)^2\\
	=&\frac{-3i\pi^2}{8(2\pi i)^2}\frac{1}{(x_{14}x_{23})^4}\iint\limits_{c-i\infty}^{c+i\infty}\dd s\dd t\frac{v^{s+t}}{u^t}\Gamma(s)^2\Gamma(t)^2\Gamma(2-s-t)^2
\end{align}
Now the integration over $s$ is done by using the following identity for an inverse Mellin transformation:
\begin{align}
	\frac{1}{2\pi i}\int\limits_{c-i\infty}^{c+i\infty}\dd s\Gamma(s)^2\Gamma(2-s-t)^2v^s&=\Gamma(2-t)^4\frac{F\left(2-t,2-t;2(2-t);1-v^{-1}\right)}{\Gamma(2(2-t))}\\
	&=\Gamma(2-t)^2\sum_{n=0}^{\infty}\frac{\Gamma(2-t+n)^2}{\Gamma(2(2+t)+n)n!}(1-v^{-1})^n
\end{align}
With this the above expression becomes:
\begin{align}
	M_4=&\frac{-3i\pi^2}{8(2\pi i)}\frac{1}{(x_{14}x_{23})^4}\sum_{n=0}^{\infty}\frac{(1-v^{-1})^n}{n!}\int\limits_{c-i\infty}^{c+i\infty}\dd t\frac{v^{t}}{u^t}\Gamma(t)^2
	\Gamma(2-t)^2\frac{\Gamma(2-t+n)^2}{\Gamma(2(2+t)+n)}
\end{align}
The last integral can now be performed by choosing the integration contour with negative real part. The only poles in this contour come from $\Gamma(t)^2$ which has poles at negative integer values and therefore the residues at these points are:
\begin{align}
	\Res_{t=m<0}&\left\{\frac{v^{t}}{u^t}\Gamma(t)^2\Gamma(2-t)^2\frac{\Gamma(2-t+n)^2}{\Gamma(2(2+t)+n)}\right\}\\
	=&2\frac{v^m}{u^m}\frac{\Gamma(2-m)^2\Gamma(2-m+n)^2}{\Gamma(4-2m+n)((-m)!)^2}\\
	&\times\left(\frac12\log\left(\frac{v}{u}\right)+\psi(1-m)-\psi(2-m)+\psi(4-2m+n)-\psi(2-m+n)\right)
\end{align}
So by using the residue theorem the final result can be written as:
\begin{align}
	M_4=&\frac{-3i\pi^2}{4}\frac{1}{(x_{12}x_{34})^4}\left(\frac{u}{v}\right)^2\sum_{n,m=0}^{\infty}\frac{(1-v^{-1})^n(u/v)^m}{n!(m!)^2}
	\frac{\Gamma(2+m)^2\Gamma(2+m+n)^2}{\Gamma(4+2m+n)}\\ 
	&\times\left(\psi(1+m)-\frac12\log\left(\frac{u}{v}\right)-\psi(2+m)+\psi(4+2m+n)-\psi(2+m+n)\right)\\
	=&-i\frac{3\pi^2}{4}\frac{1}{(x_{12}x_{34})^4}\left(\frac{u}{v}\right)^2L_0\\
	L_0=&\sum_{n,m=0}^{\infty}\frac{(1-v^{-1})^n(u/v)^m}{n!(m!)^2}
	\frac{\Gamma(2+m)^2\Gamma(2+m+n)^2}{\Gamma(4+2m+n)}\\
	&\times\left(\psi(1+m)-\frac12\log\left(\frac{u}{v}\right)-\psi(2+m)+\psi(4+2m+n)-\psi(2+m+n)\right)
\end{align}
This result has the same form as the one already obtained for AdS. The main difference is the prefactor which differs from the AdS case by a factor of $-i$.
\begin{sidewaystable}
	\section{OPE coefficients $C_{n,l}^{(1)}$ and anomalous dimensions $\gamma^{(2)}_{n,l}$}
	\begin{small}
	\begin{tabular}{|c"c|c|c|c|c|}
		\hline 
		$n \setminus l$ & $0$ & $2$ & $4$ & $6$ & $8$  \\ 
		\thickhline 
		$0$ & $\frac13$ & $\frac{22}{25}$ & $\frac{1066}{1323}$ & $\frac{1327184}{2760615}$ & $\frac{9525424}{41368327}$  \\ 
		\hline 
		1 & $-\frac{478}{735}$ & $-\frac{20834676}{22204105}$ & $-\frac{4866352}{7966035}$ & $-\frac{13064388928}{43517414655}$ & $-\frac{74513809038528}{586759598886461}$ \\ 
		\hline 
		2 & $-\frac{11507}{50820}$ & $-\frac{14977826}{60144903}$ & $-\frac{15701186862675}{107693096012359}$ & $-\frac{3223599567312}{47662730167195}$ & $-\frac{4255412689576}{154269625644225}$  \\ 
		\hline 
		3 & $-\frac{278092}{6441435}$ & $-\frac{336508153240}{8369623238427}$ & $-\frac{188831734000928}{8697998341538649}$ & $-\frac{35544998611436660416}{3705879400892737336125}$ & $-\frac{31895618623739167654400}{8426142758582793609173991}$ \\ 
		\hline
		4 & $-\frac{609121355}{97749595944}$ & $-\frac{1876719645527254875}{361741682278580800196}$ & $-\frac{428047560174227}{162496693319714325}$ & $-\frac{1480492036506063088}{1323442474090734750831}$ & $-\frac{1878623428001047331688}{4375038719519998727619125}$ \\ 
		\hline
		5 & $-\frac{2949926471}{3867166172870}$ & $-\frac{141864901726234}{242136292263474375}$ & $-\frac{110172481789609092130536}{388605383582113938238057475}$ & $-\frac{3079867939300262769856}{26424621364983338048588127}$ & $-\frac{9447239300485238393536}{216190658918542321662114375}$\\
		\hline
		6 & $-\frac{320926668671}{3835597469418000}$ & $-\frac{5223413549130238378}{86374495307685977022375}$ & $-\frac{16093403644056560278}{571606009301828598928875}$ & $-\frac{5092476438104354590143024}{452028029837026462659233553125}$ & \\
		\hline
		7 & $-\frac{96685099718536}{11396815111695209625}$ & $-\frac{14250018853552940800304}{2437030404243061563964799475}$ & $-\frac{3899967031632997246896960}{1478659162782263437208306590517}$ & & \\
		\hline
		8 & $-\frac{86292642440729}{106439390863395590000}$ & $-\frac{35586037890864258418}{66157227796855523441818705}$ & & & \\
		\hline
		9 & $-\frac{3243126578062433}{43839556849961130438180}$ & & & & \\
		\hline
	\end{tabular}\\
	\\
	\\
	\begin{tabular}{|c"c|c|c|c|c|}
		\hline
		$n\setminus l$ & $10$ & $12$ & $14$ & $16$ & $18$\\
		\thickhline
		$0$ & $\frac{3756579392}{38877680205}$ & $\frac{887679309952}{23904186508125}$ & $\frac{180181278533632}{13440572799085575}$ & $\frac{11352595412240384}{2458136592718574625}$ & $\frac{11914815905718493184}{7765103693059916828625}$\\
		\hline
		$1$ & $-\frac{2287031738624}{46782666104265}$ & $-\frac{2812332880317184}{159490781312482125}$ & $-\frac{1934835084951497555968}{319092109838455308229875}$ & $-\frac{201643456886654377984}{100314038711981137168125}$ & \\
		\hline
		$2$ &  $-\frac{2754828259858104811776}{266073433910880455050625}$ & $-\frac{14322964594211936512}{3907231863192590850225}$ & $-\frac{122082761987988318040064}{98240277844866377400537357}$  &  & \\ 
		\hline
		$3$ & $-\frac{872657853615754261504}{628980119719525125646875}$ & $-\frac{7045889987881843139158016}{14604685782786989664470705295}$ &  &  & \\
		\hline
		$4$ & $-\frac{9690798970455504915008}{62858240428956503479494975}$ & & & & \\
		\hline
	\end{tabular}
\end{small}
	\caption{OPE coefficients $C_{n,0}^{(1)}$}
	\label{tab:OPE1}
\end{sidewaystable}
\begin{sidewaystable}
		\begin{tabular}{|c"c|c|c|c|c|c|c|c|c|c|}
			\hline 
			$n \setminus l$ & $0$ & $2$ & $4$ & $6$ & $8$ & $10$ & $12$ & $14$ & $16$ & $18$\\ 
			\thickhline 
			$0$ & $\frac{5}{3}$ & $-\frac{1}{20}$ & $-\frac{1}{140}$ & $-\frac{1}{504}$ & $-\frac{1}{1320}$ & $-\frac{1}{2860}$ & $-\frac{1}{5460}$ & $-\frac{1}{9520}$ & $-\frac{1}{15504} $ & $-\frac{1}{23940}$\\
			\hline
			$1$ & $\frac{46}{15}$ & $-\frac{107}{1260}$ & $-\frac{19}{1260}$ & $-\frac{131}{27720}$ & $-\frac{301}{154440}$ &
			$-\frac{19}{20020}$ & $-\frac{4}{7735}$ & $-\frac{1493}{4883760}$ & $-\frac{313}{1627920}$  & \\
			\hline
			$2$ & $\frac{113}{28}$ & $-\frac{269}{2520}$ & $-\frac{6707}{311850}$ & $-\frac{1973}{270270}$ &
			$-\frac{3439}{1081080}$ & $-\frac{59281}{36756720}$ & $-\frac{28771}{31744440}$ & $-\frac{6703}{12209400}$  &  & \\
			\hline
			$3$ & $\frac{535}{112}$ & $-\frac{6697}{55440}$ & $-\frac{19037}{720720}$ & $-\frac{143581}{15135120}$ &
			$-\frac{175939}{40840800}$ & $-\frac{349787}{155195040}$ & $-\frac{164861}{126977760}$ &  &  & \\
			\hline
			$4$ & $\frac{745007}{138600}$ & $-\frac{880571}{6756750}$ & $-\frac{13589}{450450}$ & $-\frac{138443}{12252240}$ &
			$-\frac{2775023}{523783260}$ & $-\frac{1654313}{581981400}$  &  &  &  & \\
			\hline
			$5$ & $\frac{3176419}{540540}$ & $-\frac{148199}{1081080}$ & $-\frac{6384097}{192972780}$ &
			$-\frac{1655249}{129329200}$ & $-\frac{1433129}{232792560}$ &  &  &  &  & \\
			\hline
			$6$ & $\frac{238646701}{37837800}$ & $-\frac{12184001}{85765680}$ & $-\frac{28831919}{814773960}$ & $-\frac{102992111}{7332965640}$ &
			&  &  &  &  & \\
			\hline
			$7$ & $\frac{65520575}{9801792}$ & $-\frac{105623209}{724243520}$ & $-\frac{11557961}{310390080}$ &  &  &  &
			&  & & \\
			\hline
			$8$ & $\frac{980571811}{139675536}$ & $-\frac{51952177}{349188840}$ &  &  &  &  &  &  & & \\
			\hline
			$9$ & $\frac{2983181297}{407386980}$ & & & & & & & & & \\
			\hline
		\end{tabular}
	\caption{Anomalous dimension $\gamma_{n,l}^{(2)}$ in units of $\frac{\lambda_R^2}{256\pi^4}$}
	\label{tab:gammanl}
\end{sidewaystable}
\newpage

\bibliography{references}
\bibliographystyle{JHEP}
\end{document}